\documentclass[pra,groupedaddress,reprint,showpacs,amsmath,amssymb,nofootinbib]{revtex4-1}

\usepackage{bbold,color} 
\usepackage{graphicx}
\graphicspath{{graphics/}}

\usepackage{subfig}
\usepackage{amsmath}
\usepackage{verbatim}
\usepackage{float}
\newcommand{\comments}[1]{}
\def\({\left(}
\def\){\right)}

\def\ad{a^\dagger}
\def\d#1{#1^\dagger}

\newcommand\eq[1]{Eq.~(\ref{eq:#1})}

\newcommand\eqs[2]{Eqs.~(\ref{eq:#1}-\ref{eq:#2})} 
\newcommand\Eqs[2]{Equations~(\ref{eq:#1}-\ref{eq:#2})} 
\newcommand\fig[1]{Fig.~\ref{fig:#1}}
\newcommand\figs[2]{Figs.~\ref{fig:#1} and \ref{fig:#2}}

\newcommand\se[1]{section~\ref{sec:#1}}
\def\ket#1{\left|\,#1\,\right\rangle}

\begin{document}

\title{Generation of multipartite spin entanglement from multimode squeezed states}
\author{Niranjan Sridhar}
\email{ns4mf@virginia.edu}
\author{Olivier Pfister}
\email{opfister@virginia.edu}
\affiliation{Department of Physics, University of Virginia, Charlottesville, Virginia 22903, USA}

\date{\today}

\begin{abstract}
We investigate the systematic use of the Schwinger representation, by virtue of which two boson fields are equivalent to an effective spin, for casting multimode squeezed states into multipartite spin entangled states. The motivation for this endeavor is the following established fact: finitely squeezed, two-mode entangled states can be recast into maximally entangled bipartite spin states, irrespective of their level of squeezing (in the lossless case). This work explores the generalization of this interesting property to multipartite entanglement. While we found that the generalization of multipartite Gaussian entanglement to multipartite spin entanglement is not straightforward, there are nonetheless interesting features and entangled states to be found. Here we study a few   CV entangled states already realized experimentally, and show that some of them correspond to multipartite spin entangled states.
\end{abstract}

\maketitle

\section{Introduction}

Continuous-variable (CV) entanglement is a highly interesting and active field because it provides a new outlook on quantum information, offers rich perspectives such as massive scalability potential~\cite{Pfister2004,Menicucci2008,Pysher2011,Yokoyama2013,Roslund2013}, and can rely on the mature quantum optical experimental techniques of squeezed-state generation. In particular, our group and collaborators have discovered~\cite{Menicucci2008,Flammia2009} and begun to demonstrate~\cite{Pysher2011} massively scalable continuous-variable cluster entanglement in the quantum optical frequency comb, which opens up opportunities towards generating macroscopic entangled states for universal quantum computing.

However, the no-go theorems for all-Gaussian key quantum processes such as Bell inequality violation~\cite{Bell1987}, entanglement distillation~\cite{Eisert2002}, and quantum error correction~\cite{Niset2009} require us to seek either non-Gaussian measurements/gates on Gaussian states~\cite{Menicucci2006} or to generate non-Gaussian states directly. Here we explore the former approach by first casting Gaussian states as effective spins, by use of the Schwinger representation~\cite{Schwinger1965}. Spin measurements will then coincide with photon-number-resolving measurements~\cite{Lita2008}  in the Fock state basis, which are known to be non-Gaussian measurements since Fock states have nonpositive Wigner functions~\cite{Hudson1974}. 

Another goal is to ascertain whether the Schwinger representation would be a possible bridge from massively scalable Gaussian entanglement to massively scalable spin entanglement and, possibly, quantum simulation~\cite{Feynman1982} of entangled spin lattices.
 
Here, we present a first nontrivial step toward these goals with the theoretical discovery that quadripartite spin states which are entangled to make total spin zero states can be achieved using quadripartite Gaussian entanglement. Moreover, we discovered that the number of entangled spins and the strength of their entanglement is {\em  independent} of the strength of the nonlinear squeezing interaction, in the absence of losses. 

However, the correspondence between CV and spin entanglement is not obvious and it is not clear if its possible to generate other entangled spin states using CV entanglement and there is no straightforward map yet between CV and spin entangled states. Therefore, we attempt to develop a theoretical framework for finding underlying spin symmetries in squeezing Hamiltonians. In light of recent advances in photon-number resolving detection, we anticipate that this will be a powerful way to generate and simulate not just Qbit\footnote{We adopt here the more harmonious spelling of David Mermin's~\cite{Mermin2003}.} but multipartite high-spin entanglement. 

Note that related work has been produced by Natasha Gabay and Nicolas Menicucci~\cite{Gabay2013}, in connection with the Gaussian graph formalism developed by Menicucci~\cite{Menicucci2011,Menicucci2011a}.

The paper is organized as follows. In \se{Schwing} we review the Schwinger representation, a. k. a. the quantum Poincar\'e sphere, or quantum Stokes parametrization and illustrate its physical significance as well as its application to the example of bipartite spin entanglement using two EPR pairs. In \se{Symmetry} we look at the theoretical methods that allow us to find spin operators that are constants of motion of the squeezing Hamiltonian and can be used to define the states that they nullify. We then re\"examine the example of bipartite spin entanglement using two EPR pairs. In \se{nullifs}, we extend the results of our systematic derivations of the spin nullifiers to twin tri- and quadripartite CV states. In \se{entang}, we make use of these results to derive the corresponding spin states in these cases, and examine their entanglement. We then conclude.

\section{Schwinger representation}\label{sec:Schwing}
\subsection{Mathematical formulation}
In the Schwinger representation, two bosonic fields, say of annihilation operators $a_1$ and $a_2$, are used to define an effective spin angular momentum $\vec J$, as follows 
\begin{align}
{J}_x &= \frac{1}{2}\bigl( {a}^{\dag}_1{a}_2+a_{1}{a}^{\dag}_2\bigr)  \label{eq:jx}\\
{J}_y &= \frac{1}{2i}\bigl( {a}^{\dag}_1{a}_2-a_{1}{a}^{\dag}_2\bigr)  \label{eq:jy}\\
{J}_z &= \frac{1}{2}\bigl( {a}^{\dag}_1{a}_1-{a}^{\dag}_2{a}_2\bigr),  \label{eq:jz}
\end{align}
which can easily be shown to obey the canonical commutation relations of an angular momentum. The spin ladder operators are
\begin{align}
{J}_+ &= J_{x}+iJ_{y} = {a}^{\dag}_1{a}_2  \\
{J}_- &=  J_{x}-iJ_{y} = {a}_1{a}^{\dag}_2. 
\end{align}
Finally the spin magnitude is
\begin{align}
{J}^2 &= {J}_x^2+{J}_y^2+{J}_z^2 
= \frac{{a}^{\dag}_1{a}_1+{a}^{\dag}_2{a}_2}{2}\Bigl(\frac{{a}^{\dag}_1{a}_1+{a}^{\dag}_2{a}_2}{2}+1\Bigr) \label{eq:j2}
\end{align}
and can be shown to be a scalar operator---consistent with the total energy of the two modes of the electromagnetic field. This leads us to the physical significance of this mathematical representation.

\subsection{Physical meaning}

As just remarked, $J^{2}$ represents the total energy of the two-mode field. This begets the adoption of the Fock basis $\ket{n_{1}}_{1}\ket{n_{2}}_{2}$. Indeed, consider the action of the $J^{2}$ and $J_{z}$ operators of a two-mode number state, using \eq{jz} and \eq{j2}:
\begin{align}
J^{2} \ket{n_{1}}_{1}\ket{n_{2}}_{2}&= \frac{n_{1}+n_{2}}2\Bigl(\frac{n_{1}+n_{2}}2+1\Bigr) \ket{n_{1}}_{1}\ket{n_{2}}_{2} \\
J_{z} \ket{n_{1}}_{1}\ket{n_{2}}_{2}&= \frac{n_{1}-n_{2}}2 \ket{n_{1}}_{1}\ket{n_{2}}_{2}.
\end{align}
This proves that the two-mode Fock states are the spin eigenstates $\ket{j m}$, with
\begin{align}
j &= \frac{n_{1}+n_{2}}2 \\
m &= \frac{n_{1}-n_{2}}2. 
\end{align}
Hence the spin magnitude is the total photon number and the $z$-component of the spin is the photon number difference. The other two components of this effective spin are equally meaningful:  \eqs{jx}{jy} clearly show that $J_{x,y}$ are interference terms of fields 1 and 2, respectively in phase and in quadrature. 

Note that in this paper we will often use interchangeably for the total photon number and spin magnitude operators the SU(2) Casimir operator 
\begin{equation}
{J}_0 = \frac12({a}^{\dag}_1{a}_1+{a}_2^{\dag}{a}_2),
\end{equation}
whose eigenvalue is $j$ and which verifies ${J}^2 = {J}_0({J}_0+1)$.

\begin{figure}[h]
\centering
\includegraphics[width=\columnwidth,trim=160 130 170 120,clip]{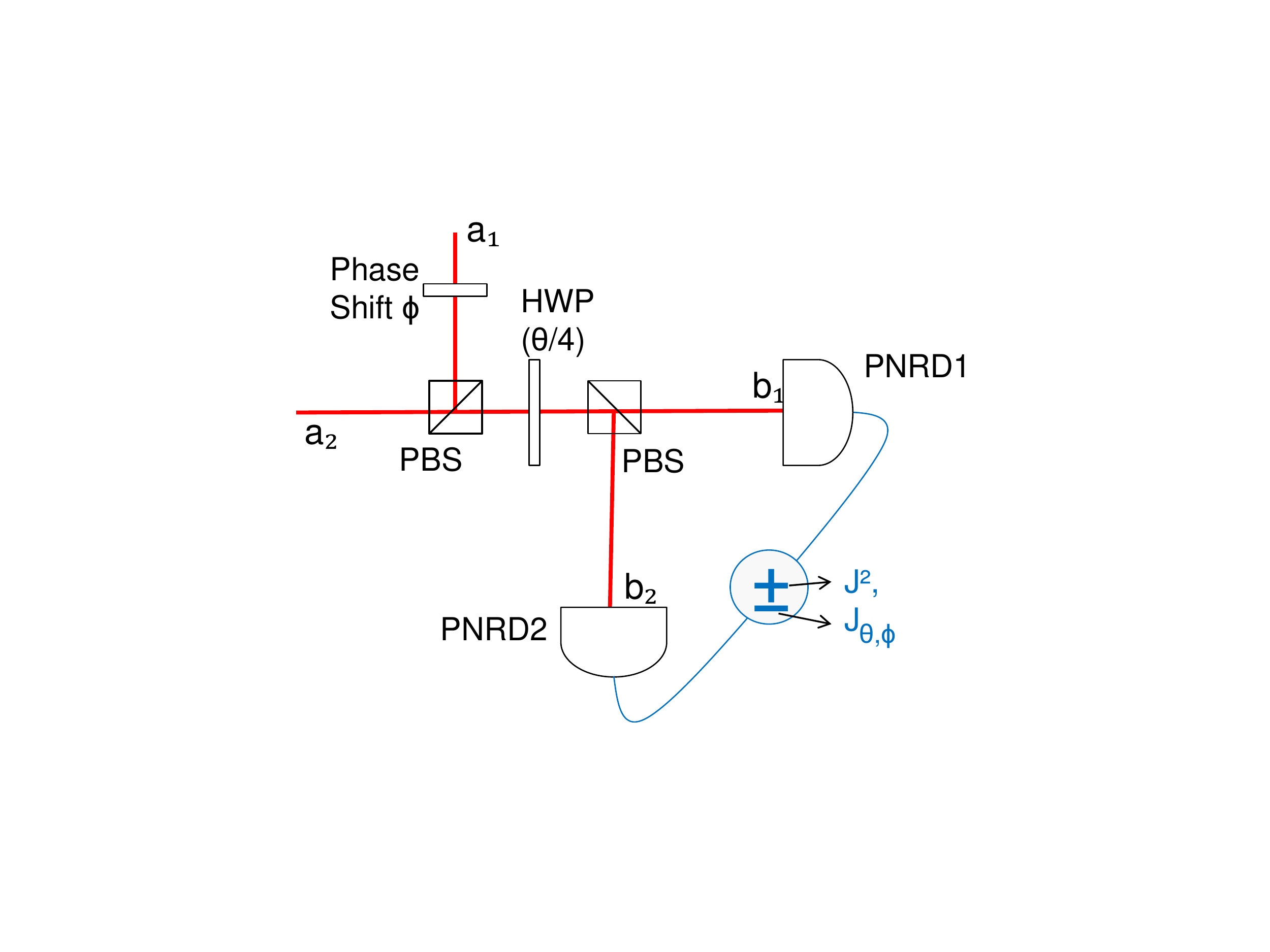}
\caption{ Arbitrary spin measurements (along direction $(\theta,\phi)$) can be performed with a phase shift $\phi$, two polarizing beamsplitters (PBS), a halfwave plate (HWP) whose axes are at $\theta/4$ rad from the PBSs, and two photon-number-resolving detectors (PNRD1,2).}
\label{fig:SpinMeasure}
\end{figure}

From these considerations, we easily deduce that measurements of the effective spin {\em along any direction} can then be made using variable beamsplitters and photon number resolving detection~\cite{Yurke1986a,Kim1998,Evans2011}, as depicted in \fig{SpinMeasure}.

The detected fields are
\begin{align}
{b}_1 &= \frac1{\sqrt{2}}\Bigl(a_{1}\cos\frac\theta2 {}+a_{2}\, e^{-i\phi}\sin\frac\theta2 \Bigr) \\
b_2 &= \frac1{\sqrt{2}}\Bigl(-a_{1}\sin \frac\theta2 {}+ a_{2}\, e^{-i\phi}\cos\frac\theta2 \Bigr) 
\end{align}
and the corresponding photon numbers are
\begin{align}
{b}^{\dagger}_1{b}_1 &= \frac{1}{2} [{N}_1\,\cos^2\frac\theta2  + {N}_2\,\sin^2\frac\theta2  \nonumber \\
&\qquad + ( {a}^{\dagger}_{1} {a_{2}}\,e^{-i\phi}+{a}_{1}a_{2}^{\dagger}\,e^{i\phi})\,\cos\frac\theta2 \sin\frac\theta2 ]  \\
{b}^{\dagger}_2{b}_2 &=\frac{1}{2} [{N}_1\,\sin^2\frac\theta2  + {N}_2\,\cos^2\frac\theta2  \nonumber \\
&\qquad -  ( {a}^{\dagger}_{1}{a_{2}}\,e^{-i\phi}+{a}_{1}a_{2}^{\dagger}\,e^{i\phi})\,\cos\frac\theta2 \sin\frac\theta2 ] \end{align}
so that
\begin{align}
N_{+}=\d b_{1}b_{1}+\d b_2b_2 &= \frac12({N}_1+{N}_2)=J_{0} 
\end{align}
and
\begin{align}
N_{-}&=\d b_{1}b_{1}-\d b_2b_2 \nonumber \\
&= \frac{{N}_1-{N}_2}2\,{\cos\theta}+\frac{a_{1}^{\dagger}{a_{2}}\,e^{-i\phi}+{a_{1}}{a_{2}}^{\dagger}\,e^{i\phi}}2\,{\sin\theta} \nonumber  \\
&= {J}_z\,\cos\theta  + ({J}_x\,\cos\phi  + {J}_y\,\sin\phi )\,\sin\theta
\end{align}
which prescribes how to set the half-wave plate $\theta$ and the phase shift $\phi$ to measure any component of the spin, together with its magnitude. 

\subsection{Previous results} 

The Schwinger representation has been widely used over the years, starting from the group theoretical modeling of interferometers by Yurke et al.\ in 1986~\cite{Yurke1986a}. 

In 2002, Bowen et al.\ experimentally demonstrated single-spin squeezing from two-mode squeezed light~\cite{Bowen2002}. 

In 2005, C. Gerry and J. Albert proposed using a beamsplitter with a Fock state input to violate Bell inequality using the Holstein-Primakoff spin representation. In this representation, {\em single-mode} Fock states correspond to $J_z$ eigenstates (as opposed to the two Qmodes in the Schwinger representation), the vacuum state $|0\rangle$ corresponds to the $|j,-j\rangle$ state, and $|n\rangle$ to $|j,-j+n\rangle$.  The photon number anticorrelation between the two output Qmodes of a beamsplitter can then be written as an entangled spin system. 

In 2011, Evans and Pfister used an original proposal by Drummond and Reid~\cite{Drummond1983,Reid2002} to show theoretically~\cite{Evans2011} that entangled spins could be used to violate the Mermin inequality~\cite{Mermin1980}. This proposal uses the  photon correlation in the two-mode squeezed state produced by parametric down-conversion (PDC) in an optical parametric amplifier (OPA) to create perfectly entangled spins of arbitrary magnitude. Indeed, two independent OPAs (labeled 1 and 2) emit a tensor product of EPR states, or SU(1,1) Perelomov coherent state~\cite{Gerry2001},
\begin{align}
\ket{EPR^2} &= \sum_{n_1=0}^\infty  \frac{\tanh^{n_1} r_1}{\cosh r_1}\ket{n_1}_{A1}\ket{n_1}_{B1} \nonumber\\
&\qquad \otimes\sum_{n_2=0}^\infty  \frac{\tanh^{n_2} r_2}{\cosh r_2}\ket{n_2}_{A2}\ket{n_2}_{B2},\label{eq:twinEPR}
\end{align}
which can be rewritten in the Schwinger representation of spins $A$ ($A1$,$A2$) and $B$ ($B1$,$B2$), to a local optical phase shift left and assuming equal squeezing parameters $r_{1}=r_{2}=r$, as a superposition of {\em maximally entangled} states of zero total spin:
\begin{equation}
\ket{EPR^2} = \sum_{j=0}^\infty \frac{\tanh^{2j} r}{\cosh^2 r}
\sum_{m=-j}^j  (-1)^{j-m}\ket{j, m}_A\ket{j, -m}_B \label{eq:twinspins}
\end{equation}
where $j=(n_{1}+n_{2})/2$ as before. Note that the recast of \eq{twinEPR} as spin eigenstates features a rather remarkable property: the entanglement amount, initially quantified by the squeezing parameter in the EPR state of \eq{twinEPR}, becomes independent of $r$ when the entangled part of the state is expressed  as an SU(2) eigenstate in \eq{twinspins}, as can be clearly seen from its rightmost sum. In \eq{twinspins}, the squeezing parameter $r$ only conditions the probability of observing a particular spin magnitude $j$, {\em not the degree of entanglement}. For each and every value of $j$, the entanglement is maximal and independent of $r$.

The aforementioned bipartite entanglement property provides us with a strong motivation for investigating connections between Gaussian and spin entanglement in the multipartite case.

Although well known in quantum optics~\cite{Wodkiewicz1985,Yurke1986a}, this interplay between the SU(1,1) and SU(2) groups is compelling and its consequences for entanglement have not yet been elucidated, to the best of our knowledge. A more general description, such as that involving the symplectic group Sp(4,$\mathbb R$)~\cite{Arvind1995}, might be useful here as it already has been for the study of Gaussian entanglement~\cite{Simon2000}, but these theoretical directions are beyond the scope of the present paper.

\section{Quantum evolution of CV- and spin fields. Bipartite case}\label{sec:Symmetry}

We now turn to the casting of CV multipartite entangled states into spin states and investigating the entanglement of the latter. We naturally start with the simplest nontrivial examples
of tripartite and quadripartite~\cite{Zaidi2008,Pysher2011} CV graph states.  

\subsection{Quantum evolution of CV, nullifiers, stabilizers, and constants of the motion}
We write the complete basis of the quadrature operators for $n$  quantum modes (``Qmodes'') as a vector $(Q,P)^T$ where $Q = ( Q_1, ..., Q_n )$, $P= (P_1, ..., P_n)$, and where $Q_j=(a_j+\ad_j)/\sqrt2$ and $P_j=i(\ad_j-a_j)/\sqrt2$, $a_j$ being the photon annihilation operator of Qmode j. We use quadratic squeezing Hamiltonians of the form 
{\begin{align}
 H&= i\hbar\kappa \sum_{j,k} (\ad_j G_{jk}\ad_k - a_jG_{jk}a_k) \\
&=  \frac{\hbar\kappa}2(Q^{T}GP+P^{T}GQ),
\end{align}}
{where $\kappa>0$ and $G$ is the H(amiltonian)-graph adjacency matrix~\cite{Zaidi2008,Menicucci2013}, whose 0 and 1 entries inform on which Qmodes of the field are subjected to a two-mode squeezing interaction. From the physical point of view, diagonalizing $G$ solves the Heisenberg-equation system~\cite{Pfister2004}
\begin{align}
\dot Q&= \kappa GQ \\
\dot P&= -\kappa GP.
\end{align}
 From the mathematical, graph theoretical point of view, diagonalizing $G$ yields the spectrum of the H-graph.} 
{Let $G_D$ be the diagonal form of $G$ and $M$ the diagonalization matrix, then}
\begin{align}
G &= M^{-1} G_D M
\end{align}
and  the ``eigenoperators''
{$Q' = MQ$ and $P' = MP$ of course verify  
\begin{align}
\dot Q'&= \kappa G_D Q' \\
\dot P'&= -\kappa G_D P' 
\end{align}
} 
which leads to the familiar result that the negative eigenvalues of $G$ imply amplitude-quadrature squeezing 
\begin{equation}
Q'_{j}(t)=Q'_{j}(0)\,e^{-|G_{Djj}|r},
\end{equation}
where $r=\kappa t$ is the squeezing parameter, and that the positive eigenvalues imply phase-quadrature squeezing
\begin{equation}
P'_{k}(t)=P'_{k}(0)\,e^{-G_{Dkk}r}.
\end{equation}
(Keep in mind that these primed Qmodes are linear superposition of the initial Qmodes $(Q,P)^{T}$). In the limit $r\gg1$, the system evolves into a simultaneous eigenstate of all the squeezed Qmodes, with eigenvalue 0. These nilpotent squeezed Qmode quadratures, a.k.a.\ {\em nullifiers}, are therefore the infinitesimal operators of the {\em stabilizers} of that same state and, more precisely, the generators of the stabilizer group~\cite{Gu2009}---the nullifiers are also the {\em variance-based entanglement witnesses}~\cite{Hyllus2006} of the same state. This Heisenberg picture for defining CV entangled states is therefore a direct analog of the stabilizer formalism used to describe Qbit entanglement~\cite{Gottesman1997}, and these squeezed (and antisqueezed) Qmodes provide a good starting point for understanding the relationship between CV- and spin entanglement.

Note finally that a zero eigenvalue of $G_D$ implies that the corresponding Qmode is a {\em constant of the motion}, commuting with the Hamiltonian. 
The measurement noise of quantum optical constants of the motion is therefore always the vacuum (or ``shot'') noise level when they evolve from an initial vacuum state.

\subsection{Quantum evolution of Schwinger spin operators}
We return to the Schwinger representation. Since all Schwinger spin operators are quadratic in field operators, we can form the time-evolved spins out the CV operators above. Of particular interest to us, in analogy with the Qbit stabilizer formalism, are the spin nullifiers. The reason for this is the following: for spin magnitudes larger than $1/2$, the unitary matrix-group representation of SU(2) is unitary but not Hermitian, unlike the Pauli group for which the Qbit stabilizers are unitary observables. For Qdits and Qmodes, the spin observables are only nullifiers, not stabilizers. (In both cases, the stabilizers are of course rotation operators and therefore exponentiated spins.)

Rewriting \eqs{jx}{jz} in terms of quadratures, we get
\begin{align}
{J}_x &=  Q_1{Q}_2+P_{1}P_2  \label{eq:jxq}\\
{J}_y &=  Q_1{P}_2-P_{1}Q_2  \label{eq:jyq}\\
{J}_z &= \frac{1}{2}\bigl( {Q}^{2}_1+{P}^{2}_1-{Q}^{2}_2-{P}^{2}_2\bigr),  \label{eq:jzq}
\end{align}

Now, If we assume $G$ to be full-rank (which covers all the interesting cases of multipartite CV entanglement), we can consider the set of all linear Qmode operators as either squeezed or antisqueezed operators. From this, we see that we can make 2 types of spin nullifiers. 

The first type is formed by a product of two squeezed Qmodes. These will be perfect nullifiers in the limit of infinite squeezing. 

The second type, however, is {\em independent of the squeezing} and is formed products of one squeezed and one antisqueezed operator. If the squeezing strengths are equal, which we'll assume throughout the rest of the paper, then these field-quadratic operators are constants of the motion. Moreover, if these products are normally ordered, then they nullify the initial vacuum state by virtue of $a\ket0=0\ket0$ and, being constants of the motion, they will also nullify the final state of the multimode squeezing Hamiltonian, whatever the value of the squeezing parameter. 

It so happens that all Schwinger spin operators defined by \eqs{jx}{jz} and \eqs{jxq}{jzq} nullify the vacuum state. Therefore we can find all spin nullifiers (i.e., stabilizers!) of any Schwinger state by systematically taking all the normally ordered products of squeezed and antisqueezed quadratures, given by \eqs{jxq}{jzq}. Again, even though these nullifiers and stabilizers are constructed out of squeezed and antisqueezed operators, they will be independent of the squeezing parameter $r$.

\begin{figure}[H]
\centering
\includegraphics[width=.3\columnwidth]{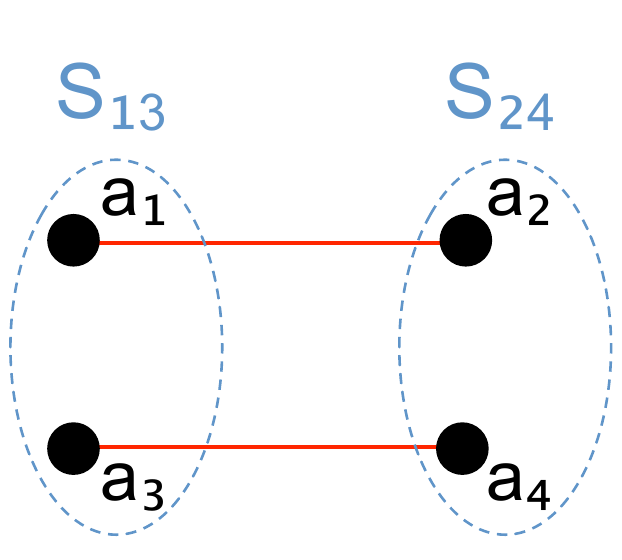}
\caption{Two sets of 2-mode squeezed states, 1-2 and 3-4, can be viewed through the Schwinger representation as 2 effective spins (blue ellipses). The black dots represent Qmodes and the red edges denote the non-zero $G_{jk}$ terms in the squeezing Hamiltonian.}
\label{fig:2Spins}
\end{figure}

Let's illustrate the above concepts in the familiar case of two 2-mode EPR pairs coupled to form two entangled spins (\fig{2Spins}), as already evoked in the Schr\"odinger picture in \eqs{twinEPR}{twinspins}.
The Hamiltonian for this system is
\begin{align}
H&=i\hbar \kappa(\ad_1 \ad_2+\ad_3 \ad_4-a_1a_2-a_3a_4) \\
&=\hbar \kappa(Q_{1}P_{2}+P_{1}Q_{2}+Q_{3}P_{4}+P_{3}Q_{4})  \\
&=\frac{\hbar \kappa}2(Q^{T}GP+P^{T}GQ) 
\end{align}
where  
\begin{align}
G&=\left(\begin{array}{cccc} 0 & 1&0&0 \\ 1 &0&0&0\\0&0&0&1\\0&0&1&0\end{array}\right),  
\end{align}
which is the adjacency matrix of the H-graph of the state in \fig{2Spins} (red edges). We shall henceforth only use the H-graph to represent the Hamiltonian instead of the G matrix. Its diagonalization yields
\begin{align}
G_D &=\left(\begin{array}{cccc} 1 & 0&0&0 \\ 0 &-1&0&0\\0&0&1&0\\0&0&0&-1\end{array}\right)
\end{align}
and
\begin{align}
M &=\left(\begin{array}{cccc} 1 & 1&0&0 \\ -1 &1&0&0\\0&0&1 & 1 \\ 0&0&-1 &1\end{array}\right).
\end{align}
Therefore, in this case, we have
\begin{align}
Q'_{1,2}(t) &= Q_1(t)\pm Q_2(t) =  ({Q}_1\pm {Q}_2)e^{\pm r}\\
P'_{1,2}(t) &= P_1(t)\pm P_2(t) = (P_1\pm P_2)e^{\mp r}\\
Q'_{3,4}(t) &= Q_3(t)\pm Q_4(t) =  ({Q}_3\pm {Q}_4)e^{\pm r}\\
P'_{3,4}(t) &= P_3(t)\pm P_4(t) = (P_3\pm P_4)e^{\mp r}.
\end{align}
Therefore, we should have $4\times 4=16$ linearly independent quadratic (spin) constants of the motion, e.g., $Q'_{1}Q'_{2}$, $Q'_{1}P'_{1}$, $P'_{2}Q'_{4}$, etc. Out of these, 6 can be defined according to \eqs{jxq}{jzq}, i.e.\ are squeezing-independent spin nullifiers for a vacuum initial state (which we assume in the rest of the paper). These are
\begin{align}
&(P_1+P_2)(P_1-P_2)+({Q}_1+{Q}_2)({Q}_1-{Q}_2) \nonumber \\
&+ (P_3+P_4)(P_3-P_4)+(Q_3+Q_4)(Q_3-Q_4) \nonumber \\
=& 4(J_{013}-4J_{024}) = \mathbb0  \\
&(P_1+P_2)(P_1-P_2)+({Q}_1+{Q}_2)({Q}_1-{Q}_2) \nonumber \\
& - (P_3+P_4)(P_3-P_4)-(Q_3+Q_4)(Q_3-Q_4)  \nonumber \\
=& 4(J_{z13}-4J_{z24}) = \mathbb0  \\
&(P_1+P_2)(P_3-P_4)+(P_1-P_2)(P_3+P_4) \nonumber \\
& + ({Q}_1+{Q}_2)(Q_3-Q_4)+({Q}_1-{Q}_2)(Q_3+Q_4) \nonumber \\
=& 4(J_{x13}-4J_{x24}) = \mathbb0  \\
&(P_1+P_2)(Q_3+Q_4)+(P_1-P_2)(Q_3-Q_4) \nonumber \\
& - ({Q}_1+{Q}_2)(P_3+P_4)-({Q}_1-{Q}_2)(P_3-P_4) \nonumber \\
=& 4(J_{y13}+4J_{y24}) = \mathbb0
\end{align}
which correspond to the spin definition of \fig{2Spins}. It can be verified that their eigenstate does indeed have the form of \eq{twinspins}---with the proper Qmode-labeling conventions, and to local optical phase shifts left~\cite{Evans2011}. Indeed, the 4 nullifiers can then be written as $J_k=J_{k13}+J_{k42} \quad \forall k\in \{x,y,z\}$ and $J_{013}-J_{042}$, which nullify the maximally entangled state of zero total spin of \eq{twinspins}.

The last two spin nullifiers are 
\begin{align}
&(P_1+P_2)(P_3-P_4)-(P_1-P_2)(P_3+P_4) \nonumber \\
& + ({Q}_1+{Q}_2)(Q_3-Q_4)-({Q}_1-{Q}_2)(Q_3+Q_4) \nonumber \\
=& 4(J_{x23}-4J_{x14}) = \mathbb0  \\
&(P_1+P_2)(Q_3+Q_4)-(P_1-P_2)(Q_3-Q_4) \nonumber \\
& - ({Q}_1+{Q}_2)(P_3+P_4)+({Q}_1-{Q}_2)(P_3-P_4) \nonumber \\
=& 4(J_{y23}+4J_{y14})  = \mathbb0
\end{align} 
However, these operators pertain to  spin pairings that are different from \fig{2Spins}, gathering Qmodes 1-4 and 2-3 instead of 1-3 and 2-4. 

The remaining 10 constants of motion cannot be written as Schwinger spin operators. They are not normally ordered, i.e.\ contain photon-number non-conserving operators such as $\ad \ad +aa$. Therefore these operators, even though constants of the motion, are not nullifiers since they do not nullify the vacuum state. 

One might be tempted to check if any of the squeezed quadratic operators such as $(P_1+P_2)({Q}_1-{Q}_2)$ form spin operators that would not be constants of motion but would be nullifiers in the limit $r\gg1$. However, one can easily check that all the remaining possible spin operators are linear combinations of squeezed and antisqueezed operators and are therefore not nullifiers. 

The next step is to apply this approach to more complicated CV multipartite entangled states. At this point, we must make clear that, even though we do conduct a systematic search for state stabilizers in the paper, we have left out (for now) the far-reaching considerations of Qdit-stabilizer groups and cluster-state definition and characterization, narrowing our scope to simply determining whether (and how) multipartite-entangled CV states can be mapped onto multipartite-entangled spin states.

We will show that this method of characterizing the spin state generated by a given quadratic Hamiltonian is analytically easy, especially since methods for finding CV squeezed and antisqueezed operators are well known. Moreover, the converse process of finding a Hamiltonian to generate any desired spin state is non-trivial, thus our approach may be quite beneficial if interesting spin states are found. Here we shall only focus on the study of states with up to 4 spins, out of multipartite CV states which have been realized experimentally.

\section{Multipartite spin nullifiers}\label{sec:nullifs}

In this section, we present the results of our systematic derivations of the spin nullifiers for twin tri- and quadripartite CV states. 

\subsection{Three-spin nullifiers}

By analogy with \fig{2Spins}, we choose to examine the twin three-Qmode arrangements illustrated in \figs{3Spins}{3SpinsGHZ}: we can have a chain with 2 interactions, say 1-2 and 2-3, or we can have a triangle with 3 interactions, 1-2, 2-3 and 3-1. The CV nullifiers for the two systems will be different and hence the spin operators and constants of motion are expected to be different as well.

\subsubsection{Three-Qmode chain} 
\begin{figure}[h]
\includegraphics[width=.5\columnwidth]{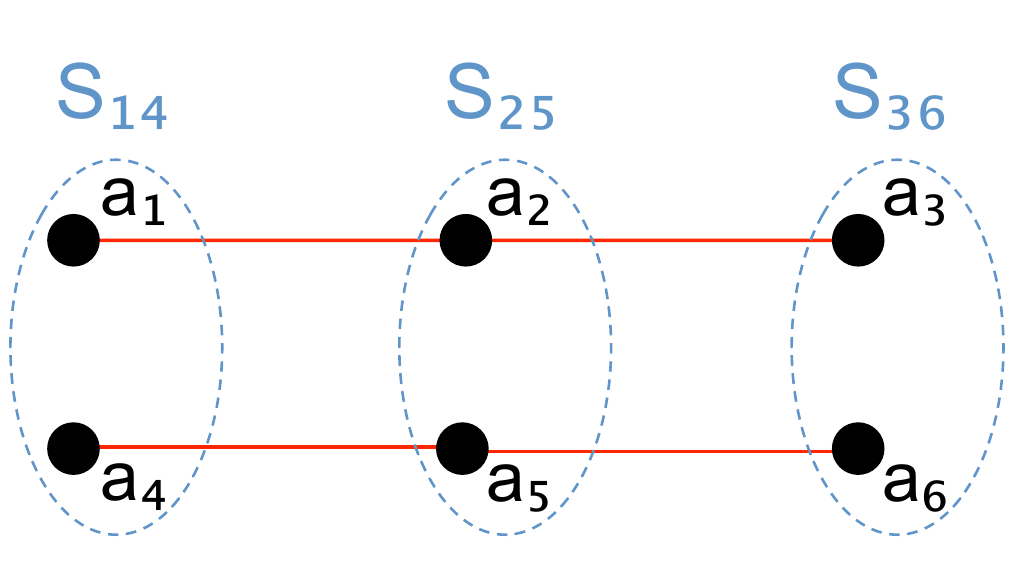}
\caption{Two sets of 3 Qmodes, 1-3 and 4-6, cast as 3 effective spins (blue ellipses).}
\label{fig:3Spins}
\end{figure}
We can easily find the spin constants of motion of the state of \fig{3Spins} using the methods developed above. The CV nullifiers are readily obtained
\begin{align}
Q'_{2}(r)&=({Q}_1-\sqrt{2}{Q}_2+Q_3)\,e^{-\sqrt2 r} \\ 
P'_{1}(r)&=(P_1+\sqrt{2}P_2+P_3)\,e^{-\sqrt2 r} \\
Q'_{5}(r)&=({Q}_4-\sqrt{2}{Q}_5+Q_6)\,e^{-\sqrt2 r} \\ 
P'_{4}(r)&=(P_4+\sqrt{2}P_5+P_6)\,e^{-\sqrt2 r} 
\end{align}
from which we find that there are 10 spin constants of motion 
\begin{align}
J_{014}+J_{036}-J_{025}+\mathbb1 &= \mathbb0  \label{eq:3a0}\\
J_{z14}+J_{z36}-J_{z25} &= \mathbb0  \label{eq:3az}\\
J_{x14}+J_{x36}-J_{x25} &= \mathbb0  \label{eq:3ax}\\
J_{y14}+J_{y36}+J_{y25} &= \mathbb0  \label{eq:3ay}\\
J_{x16}+J_{x34}-J_{x25} &= \mathbb0  \\
J_{y16}+J_{y34}+J_{y25} &= \mathbb0  \\
J_{x15}+J_{x35}-J_{x24}-J_{x26} &= \mathbb0  \\
J_{y15}+J_{y35}+J_{y24}+J_{y26} &= \mathbb0  \\
J_{014}+J_{036}-J_{x13}-J_{x46}+\mathbb1 &= \mathbb0  \\ 
J_{z14}+J_{z36}-J_{x13}+J_{x46} &= \mathbb0
\end{align}
The existence of operators which mix different spin definitions is something we see again in this case. While a nontrivial property, it is not consistent with the spin definitions adopted in \fig{3Spins} and we will therefore postpone investigating it. We will only consider operators that conform to one specific definition of spins, in this case 1-4, 2-5, 3-6, as in \fig{3Spins}. The 4 spin nullifiers in this case are given in \eqs{3a0}{3ay}.

Another interesting point is that the number of nullifiers has not increased, even though we added one spin.

\subsubsection{Thee-Qmode ring: CVGHZ state} \label{sec:ghz}

\begin{figure}[h]
\centering
\includegraphics[width=.35\columnwidth]{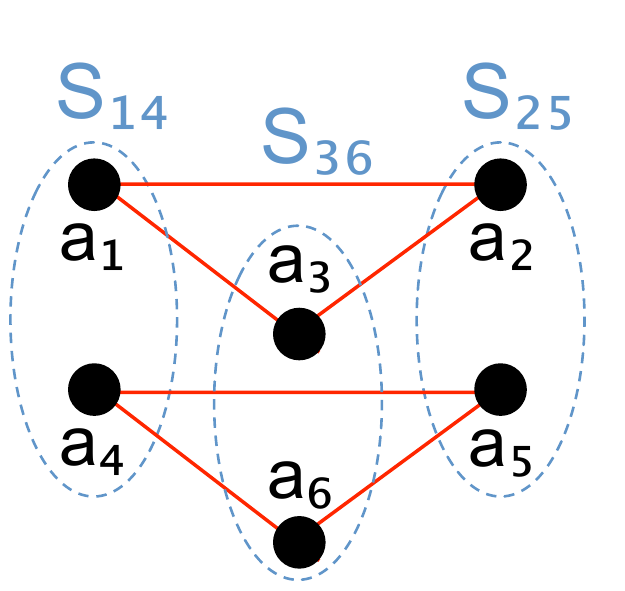}
\caption{Two sets of 3 Qmodes, 1-3 and 4-6, as 3 effective spins (blue ellipses).}
\label{fig:3SpinsGHZ}
\end{figure}

It is well known that a complete H-graph will yield a GHZ state~\cite{Pfister2004}. We are therefore naturally curious about the spin state of \fig{3SpinsGHZ}.
The CVGHZ nullifiers are well known~\cite{Braunstein2003a,Pfister2004}:
\begin{align}
P'_{1}(r)&=(P_1+P_2+P_3)\,e^{-2r} \\ 
Q'_{2}(r)&=({Q}_1-{Q}_2)\,e^{-r} \\ 
Q'_{3}(r)&=({Q}_2-Q_3)\,e^{-r} \\ 
P'_{4}(r)&=(P_4+P_5+P_6)\,e^{-2r} \\ 
Q'_{5}(r)&=({Q}_4-{Q}_5)\,e^{-r} \\ 
Q'_{6}(r)&=({Q}_5-Q_6)\,e^{-r} 
\end{align}
and from these we find the spin constants of motion of this system to be
\begin{align}
J_{y12}+J_{y23}+J_{y31} &= \mathbb0   \\
J_{y45}+J_{y56}+J_{y64} &= \mathbb0   \\
J_{y16}-J_{y14}+J_{y24}-J_{y26} &= \mathbb0   \\
J_{y15}-J_{y14}+J_{y34}-J_{y35} &= \mathbb0   \\
J_{y15}-J_{y14}+J_{y24}-J_{y25} &= \mathbb0   \\
J_{y16}-J_{y14}+J_{y34}-J_{y36} &= \mathbb0   \\
J_{y14}+J_{y15}+J_{y16}+J_{y24}\qquad\qquad\qquad &\nonumber \\
\qquad +J_{y25}+J_{y26}+J_{y34}+J_{y35}+J_{y36} &= \mathbb0
\end{align}
Interestingly enough, {\em none} of the above nullifiers is consistent with the nonoverlapping spin pairings in the spin definitions of \fig{3SpinsGHZ}! Therefore it becomes difficult to conceive of merely characterizing a three-spin system, let alone quantifying any spin entanglement, from these operators. (Recall that each spin correspond to a well-defined Qmode pair on which interference and photon-number measurements are made.)  This particular arrangement of Qmodes in \fig{3SpinsGHZ} thus seems to thoroughly defeat our approach, an interesting conclusion that stems solely from the Heisenberg viewpoint. Such is not the case, however, for the spin graph of \fig{3Spins}. Before we investigate the spin state associated with it, we derive the spin nullifiers for some 4-spin cases.

\subsection{Four-spin nullifiers}

\subsubsection{Four-Qmode chain and ring CV states} \label{sec:4}

\begin{figure}[h]
\subfloat[Chain H-graph, square cluster state, $H^{(0)}$]{\label{4SpinsChain}\includegraphics[width=.5\columnwidth]{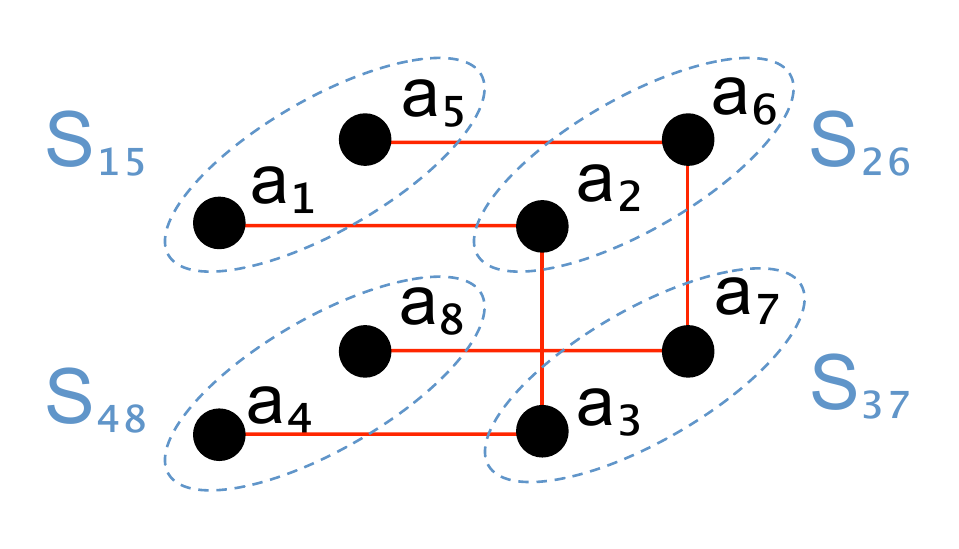}} \\
\subfloat[Square H-graph, square cluster state, $H^{(1)}$]{\label{4SpinsCluster}\includegraphics[width=.5\columnwidth]{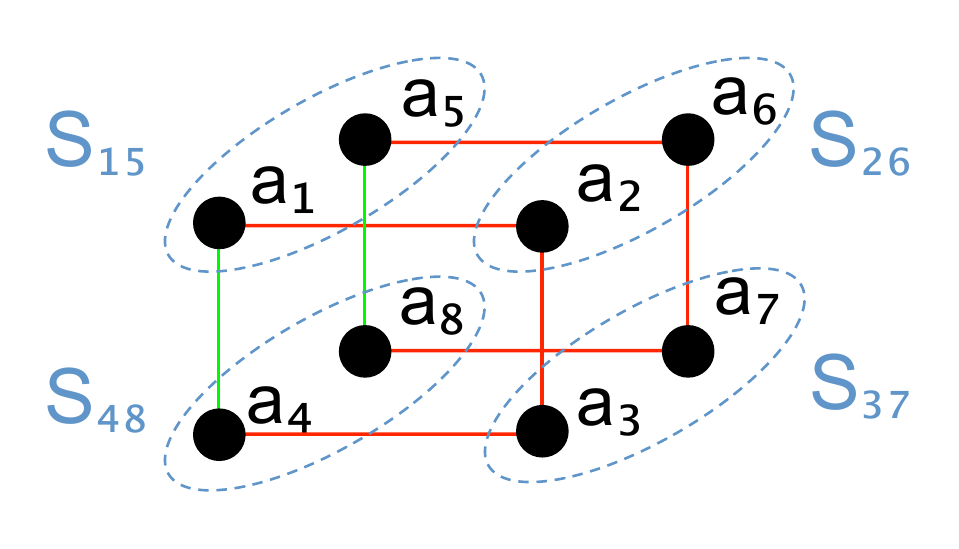}} \\
\subfloat[Ring H-graph, $H^{(2)}$]{\label{2SqSpins}\includegraphics[width=.5\columnwidth]{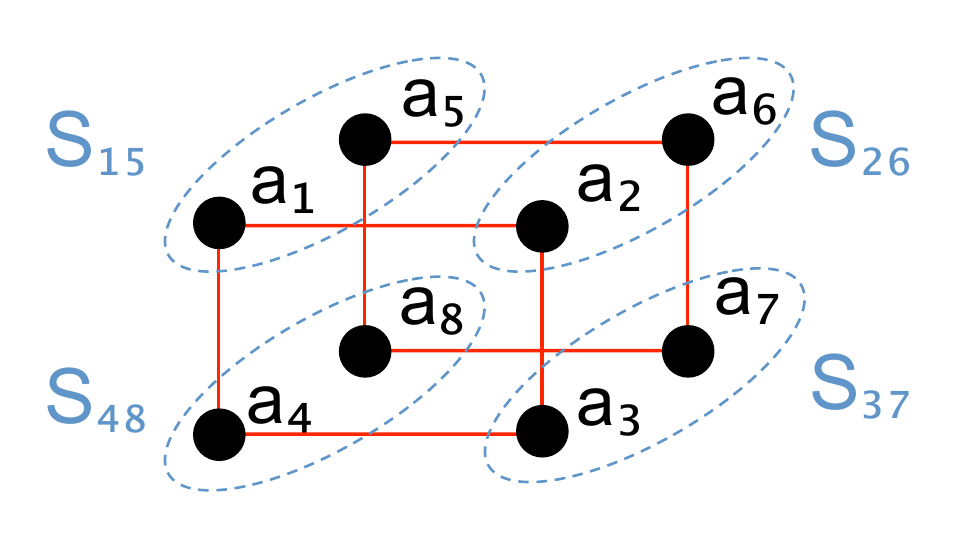}}
\caption{The 4-mode Hamiltonians studied in this paper. The green edges denote a change sign in the corresponding two-mode squeezing term with respect to the red edges.}
\label{fig:4Spins}
\end{figure}

We will not treat the 4-Qmode chain and the ring separately as these two are specific cases of a more general CV cluster state~\cite{Zhang2006}, as we first recall. (Note that it is the {\em complete} H-graph, not the {\em ring} H-graph, that gives a CVGHZ state. As is well known, these are the same for up to tripartite entanglement but become different for quadripartite and larger systems.) All the cases  are depicted in \fig{4Spins}. 

The Hamiltonian for the 4-Qmode chain [\fig{4Spins}(a)] is 
\begin{align}
H^{(0)}=i\hbar\kappa({a}^{\dagger}_1{a}^{\dagger}_2+{a}^{\dagger}_2{a}^{\dagger}_3+ {a}^{\dagger}_3{a}^{\dagger}_4) +H.c. 
\label{eq:h0}
\end{align}
and was implemented in our laboratory~\cite{Pysher2011} and shown to generate a quadripartite entangled CV cluster state, of nullifiers
\begin{align}
\label{eq:q+}
Q'_3(r) & = \bigl[\left(Q_1 - Q_4\right) + \Phi\left(Q_3 - Q_2\right)\bigr] e^{-r\Phi}\\
\label{eq:p+}
P'_1(r) & = \bigl[\left(P_1 + P_4\right) + \Phi\left(P_3 + P_2\right)\bigr] e^{-r\Phi}\\
\label{eq:q-}
Q'_2(r) & = \bigl[\Phi\left(Q_1 + Q_4\right) - \left(Q_3 + Q_2\right)\bigr] e^{-\frac r\Phi}\\
\label{eq:p-}
P'_4(r) & = \bigl[\Phi\left(P_1 - P_4\right) - \left(P_3 - P_2\right)\bigr] e^{-\frac r\Phi}
\end{align}
where $\Phi=(\sqrt5+1)/2$ is the golden ratio. These nullifiers can be shown to be equivalent, in the limit $r\gg1$ and to local optical phase shifts left, to the nullifiers of a ring (or ``square'') Qmode cluster-state ~\cite{Pysher2011}. Note also that a more general description of CV graph states in the presence of finite squeezing has since been expounded~\cite{Menicucci2011,Menicucci2011a}.

However, a Qmode square cluster state can also be generated by the Hamiltonian of \fig{4Spins}(b)~\cite{Zaidi2008,Pysher2011}
\begin{align}
H^{(1)}&= i\hbar\kappa ({a}^{\dagger}_1{a}^{\dagger}_2+{a}^{\dagger}_2{a}^{\dagger}_3+ {a}^{\dagger}_3{a}^{\dagger}_4-{a}^{\dagger}_4{a}^{\dagger}_1) +H.c. 
\label{eq:h1}
\end{align}
The solutions of the Heisenberg equations of motion for \eq{h1} are
\begin{eqnarray}
Q'_{2}(r)&=\left(Q_1 + Q_2 -\sqrt 2\ Q_4\right)\, e^{-r\sqrt 2}\\ 
Q'_{3}(r)&=\left(Q_1 - Q_2 + \sqrt 2\ Q_3\right)\, e^{-r\sqrt 2}\\ 
P'_{1}(r)&=\left(P_1 + P_2 + \sqrt 2\ P_4 \right)\, e^{-r\sqrt 2}\\ 
P'_{4}(r)&=\left(P_1 - P_2 -\sqrt 2\ P_3\right)\, e^{-r\sqrt 2}. 
\end{eqnarray}
Like \eqs{q+}{p-}, these squeezed operators exactly coincide with the same nullifiers of a square cluster state, to local phase shifts left and in the limit $r\gg1$.

Finally, we consider the Hamiltonian of \fig{4Spins}(c), which can be viewed as a two-mode squeezed state, each Qmode of which (1,2) being mixed with a vacuum mode (3,4) on a balanced beamsplitter. The resulting 4-Qmode Hamiltonian is
\begin{align}
H^{(2)}&= U^{\dagger}_{13}U^{\dagger}_{24}(i\hbar\kappa {a}^{\dagger}_1 {a}^{\dagger}_2 + H.c)U_{13}U_{24} \nonumber \\
&= \frac{i\hbar\kappa}{2} ({a}^{\dagger}_1+{a}^{\dagger}_3)({a}^{\dagger}_2+{a}^{\dagger}_4) +H.c. \nonumber \\
&= \frac{i\hbar\kappa}{2} ({a}^{\dagger}_1{a}^{\dagger}_2+{a}^{\dagger}_3{a}^{\dagger}_4+ {a}^{\dagger}_3{a}^{\dagger}_2+{a}^{\dagger}_1{a}^{\dagger}_4) +H.c. 
\label{eq:h2}
\end{align} 
where $U_{kl}=\exp[-\frac{\pi}4(\ad_{k}a_{l}+\ad_{l}a_{k})]$.

It is important to note that $H^{(0)}$, $H^{(1)}$ and $H^{(2)}$ only differ in the term of the 1-4 interaction, which is respectively zero [\eq{h0}], of opposite sign [\eq{h1}], and of the same sign [\eq{h2}] as the other terms, as is also clear from \fig{4Spins}. 

In $H^{(0)}$, the 1-4 interaction is absent. In $H^{(1)}$, the relative sign difference corresponds to having 3 nonlinear parametric downconverting interactions and 1 upconverting interaction, while in $H^{(2)}$ all 4 interactions are downconverting ones. We will see soon that the photon number correlations are similar in the 3 Hamiltonians, however, the field correlations are different. As a result, while $H^{(0)}$ and $H^{(1)}$ make quadripartite CV cluster states, $H^{(2)}$ does not ($G$ is not full-rank in the latter case). 

To describe all cases of \fig{4Spins} in the most general fashion, we therefore consider the following 8-mode Hamiltonian
\begin{align}
H_{1-4}&= i\hbar \kappa (G_{12}{a}^{\dagger}_1{a}^{\dagger}_2+G_{23}{a}^{\dagger}_2{a}^{\dagger}_3 \nonumber \\
&\qquad + G_{34}{a}^{\dagger}_3{a}^{\dagger}_4+G_{14}{a}^{\dagger}_4{a}^{\dagger}_1) +H.c. 
\label{eq:14}\\
H_{5-8}&= i\hbar \kappa (G_{56}{a}^{\dagger}_5{a}^{\dagger}_6+G_{67}{a}^{\dagger}_6{a}^{\dagger}_7 \nonumber \\
& \qquad + G_{78}{a}^{\dagger}_7{a}^{\dagger}_8+G_{58}{a}^{\dagger}_8{a}^{\dagger}_5) +H.c. 
\label{eq:58}\\
H_{1-8} &= H_{1-4}+H_{5-8}, 
\label{eq:Hgeneral}
\end{align} 
where $G_{ij}=\pm1,0$. We also decide, still in accord with \fig{4Spins}, on the specific choice of Qmode pairings such that the spins are made up of Qmode pairs (15), (26), (37), and (48), only. We now derive the spin nullifiers that pertain to this definition of spins.

Two constants of the motion can be deduced intuitively from inspection of the two-photon emission processes in \eq{14}, where we can easily see that when this Hamiltonian acts on the vacuum, pairs of photons are emitted or annihilated that involve the Qmode pairs (12), (23), (34), and (41). From this we  predict that 
\begin{equation}
(N_1+N_3)-(N_2+N_4)=\mathbb0,
\end{equation}
which can be proven easily:
\begin{align}
[N_1 + N_3 - N_2 - N_4,H ] &= [N_1+N_3,H]-[N_2+N_4,H] \nonumber \\
&=H-H=\mathbb 0,
\end{align}
hence the operator $N_1 + N_3 - N_2 - N_4$ is a constant of the motion. Since this operator is a nullifier of initial (vacuum) state, it also nullifies the final state. The same holds for \eq{58}, with
\begin{equation}
(N_5+N_7)-(N_6+N_8)=\mathbb0.
\end{equation}
Considering now the total Hamiltonian of \eq{Hgeneral}, we deduce the following two nullifiers 
\begin{align}
&(N_1+N_3-N_2-N_4)+(N_5+N_7-N_6-N_8) \nonumber \\
&\quad \qquad =(J_{015}+J_{037})-(J_{026}+J_{048}) =\mathbb 0\\
&(N_1+N_3-N_2-N_4)-(N_5+N_7-N_6-N_8) \nonumber \\
& \quad \qquad =(J_{z15}+J_{z37})-(J_{z26}+J_{z48})=\mathbb 0,  
\end{align}
which can also be derived from products of squeezed and antisqueezed linear operators. 

It is reasonable to take all the H-graphs paired to make spin graphs in \fig{4Spins} identical, as is implicit in \figs{2Spins}{4Spins}. From this we can find the following two nullifiers \begin{align}
(J_{x15}+J_{x37})-(J_{x26}+J_{x48})&=\mathbb 0 \\
(J_{y15}+J_{y37})+(J_{y26}+J_{y48}) &=\mathbb 0.
\end{align}
This again can be easily verified, e.g., for the former, 
\begin{align}
[J_{x15}&+J_{x37}-J_{x26}-J_{x48}, H_{1-8} ] \nonumber \\
&=i\hbar \kappa[G_{12}a^{\dagger}_5a^{\dagger}_2+G_{34}a^{\dagger}_7a^{\dagger}_4 + G_{23}a^{\dagger}_7a^{\dagger}_2+G_{14}a^{\dagger}_5a^{\dagger}_4  \nonumber \\
& \qquad -G_{12}a^{\dagger}_1a^{\dagger}_6- G_{34}a^{\dagger}_3a^{\dagger}_8- G_{23}a^{\dagger}_3a^{\dagger}_6 -G_{14}a^{\dagger}_1a^{\dagger}_8
\nonumber \\
&\qquad - (G_{56}a^{\dagger}_5a^{\dagger}_2+G_{78}a^{\dagger}_7a^{\dagger}_4 + G_{67}a^{\dagger}_7a^{\dagger}_2+G_{58}a^{\dagger}_5a^{\dagger}_4  \nonumber \\
&\qquad - G_{56}a^{\dagger}_1a^{\dagger}_6- G_{78}a^{\dagger}_3a^{\dagger}_8- G_{67}a^{\dagger}_3a^{\dagger}_6 -G_{58}a^{\dagger}_1a^{\dagger}_8)]
\nonumber \\
\end{align}
which is zero iff $G_{12}=G_{56}$, $G_{23}=G_{67}$, $G_{34}=G_{78}$, and $G_{14}=G_{58}$, i.e., iff the 1-4 H-graph is identical to the 5-8 H-graph. The commutation of $J_{y15}+J_{y37}+J_{y26}+J_{y48}$ with H follows similarly. 

Note that no constraint has yet been placed on the relative interactions strengths within a square, and these 4 nullifiers are therefore valid for all three Hamiltonians we discussed earlier. These nullifiers are also similar to the highly symmetric ones we had for the two spins case. Finally, we have also shown, using the exhaustive approach outlined in \se{Symmetry} and \se{nullifs} (finding squeezed and antisqueezed operators and combining them to form operators that are invariant under the Hamiltonian), that there are no other nullifiers to be found pertaining to these particular spin definitions, even though one can find 6 other SU(2) nullifiers pertaining to different spin pairings.

We can make the 4 nullifiers more symmetric by inessential adjustments, namely by exchanging the Qmodes in spins 26 and 48, which become 62 and 84 respectively, and by optically phase-shifting Qmodes 6 and 8 by $\pi$, which yields the following 4 nullifiers
\begin{align}
J_{0}=J_{015}+J_{037}-J_{062}-J_{084}=\mathbb 0, \label{eq:j0}\\
J_z=J_{z15}+J_{z37}+J_{z62}+J_{z84}=\mathbb 0, \label{eq:Jx}\\ 
J_x=J_{x15} +J_{x37}+J_{x62}+J_{x84}=\mathbb 0,\label{eq:Jy} \\
J_y=J_{y15}+J_{y37}+J_{y62}+J_{y84}=\mathbb 0. \label{eq:Jz}
\end{align}
It is remarkable, and worth repeating here, that these nullifiers hold irrespective of the relative signs of the interaction terms in the Hamiltonian of \eq{Hgeneral}.

Moreover, \Eqs{Jx}{Jz} define the components of a ``total spin'' $\vec J=\vec J_{15}+\vec J_{37}+\vec J_{62}+\vec J_{84}$. Indeed, we can show that
\begin{description}
\item[\em (i)] $[J_0,J_i]=0, \quad \forall  i={x,y,z}$.
\item[\em (ii)] $[J_k,J_l] = \epsilon_{klm} iJ_m, \quad \forall  {k,l,m}={x,y,z}$.
\end{description}
A few important points: because of {\em (i)}, we can measure $J_{0}$ simultaneously to any component, as was already the case in \fig{SpinMeasure}. This means that not only are $J_z,J_x,J_y$ nullifiers of the  state, they must also be nullifiers of any state that is post-selected by a measurement of $J_0$. Also, it is clear from \eqs{Jx}{Jz} that for each value of $J_0$, the state we seek must have zero total spin: $|0,0\rangle$. These are useful findings as we embark on finding the corresponding quantum spin state in the next section.

\subsubsection{Four-Qmode GHZ state}
\begin{figure}[h]
\includegraphics[width=.5\columnwidth]{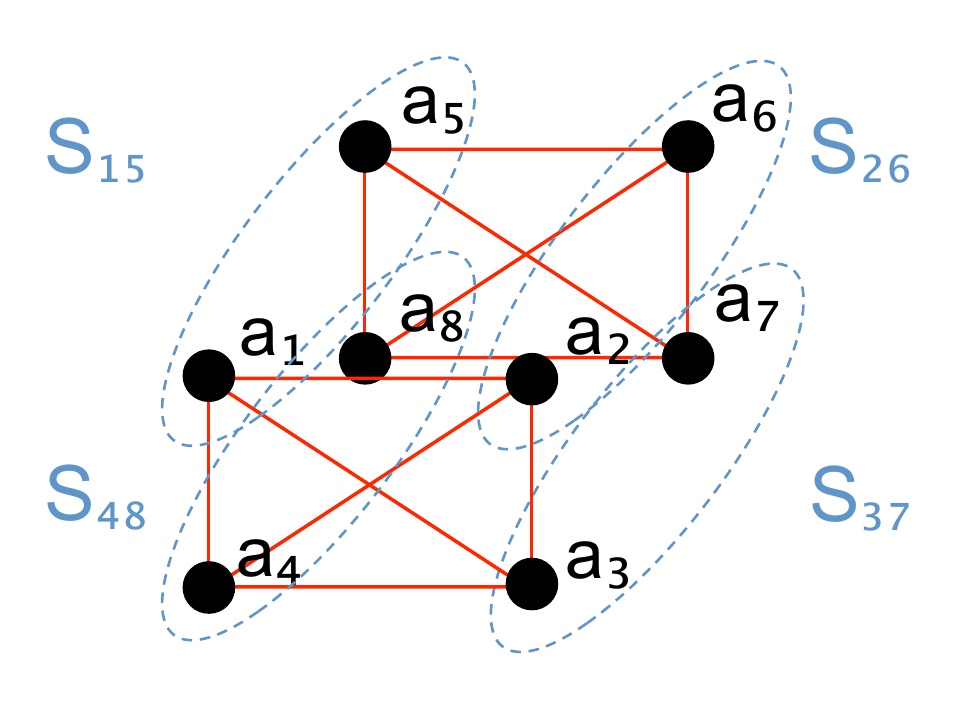}
\caption{Two 4-Qmode GHZ states, 1-4 and 5-8, paired as 4 Schwinger spins (blue ellipses).}
\label{fig:4SpinsGHZ}
\end{figure}

Before turning to the explicit expression of the spin states, we consider the same GHZ construction as in \se{ghz}, illustrated in \fig{4SpinsGHZ}. 

As mentioned earlier, the Hamiltonian is that of a complete H-graph
\begin{align}
H^{(3)} &= i\hbar\kappa \Bigl(\sum_{i<j\in[1,4]}\ad_i \ad_j +\sum_{i<j\in[5,8]}\ad_i \ad_j\Bigr)+ H.c.
\end{align}
Following again the same procedure to find the nullifiers, we found, exactly as in the 3-spin case in \se{ghz}, that there are no nullifiers that pertain the specific 4-spin definitions of \fig{4SpinsGHZ}. All the spin constants of motion that we derived mix pairings of Qmodes and therefore we cannot choose any definition of spins for which any of these operators will be applicable.

\section{Multipartite spin entanglement}\label{sec:entang}

We now turn to using the nullifiers that we have derived in the previous sections to derive the analytic expression of the corresponding spin state, in order to attempt to identify if these  spin states are multipartite entangled and, if so, to attempt to determine the nature of the entanglement. 

\subsection{Derivation of the spin state for Qbits}

Writing the full state for all photon numbers, i.e., for all values of the spin magnitudes, is an arduous and tedious task that we will not present in this paper. Here we will make use instead of the post-selection property mentioned at the end of \se{4}: indeed, we have shown that measuring the total photon number of each Qmode, i.e.\ the Casimir operator, or magnitude, of each spin ($j_{15}$, $j_{62}$, $j_{37}$, $j_{84}$) is doable simultaneously with any other measurement of the spin components that are relevant to the particular quantum process involved (e.g., Bell inequality measurement, quantum cryptography\ldots) We therefore select the simplest interesting case of 4 Qbits, i.e., $j_{15}=j_{37}=j_{62}=j_{84}={1}/{2}$, which satisfies \eq{j0}. Using this postselected substate we will prove simply that a multipartite entangled spin state is created by the Hamiltonian of \eq{h2}.

We can generate this state by decreasing the interaction strength such that we can approximate the state as the truncated expansion of the propagator on the vacuum. We do this for the Hamiltonian of \eq{Hgeneral} to get 
\begin{align}
\ket{\psi_f} &= e^{-\frac{i}{\hbar}Ht}\ket{\psi_i} \nonumber\\
&= \exp[r \{(G_{12}{a}^{\dagger}_1{a}^{\dagger}_2+G_{23}{a}^{\dagger}_2{a}^{\dagger}_3 \nonumber \\
&\qquad\quad + G_{34}{a}^{\dagger}_3{a}^{\dagger}_4 +G_{14}{a}^{\dagger}_4{a}^{\dagger}_1) -H.c. \nonumber \\ 
& \qquad\quad +(G_{12}{a}^{\dagger}_5{a}^{\dagger}_6+G_{23}{a}^{\dagger}_6{a}^{\dagger}_7 \nonumber \\
&\qquad\quad + G_{34} {a}^{\dagger}_7{a}^{\dagger}_8+G_{14}{a}^{\dagger}_8{a}^{\dagger}_5) -H.c.\}] \ket{0} \nonumber \\
&= 1 + r (...)  \nonumber \\
& \qquad + r^2(G_{12}G_{34}({a}^{\dagger}_1{a}^{\dagger}_2{a}^{\dagger}_3{a}^{\dagger}_4 + {a}^{\dagger}_1{a}^{\dagger}_2{a}^{\dagger}_7{a}^{\dagger}_8 \nonumber \\
&\qquad \qquad+{a}^{\dagger}_5{a}^{\dagger}_6{a}^{\dagger}_3{a}^{\dagger}_4 +{a}^{\dagger}_5{a}^{\dagger}_6{a}^{\dagger}_7{a}^{\dagger}_8) \nonumber \\   
& \qquad \quad +G_{23}G_{14}({a}^{\dagger}_1{a}^{\dagger}_2{a}^{\dagger}_3{a}^{\dagger}_4 +{a}^{\dagger}_1{a}^{\dagger}_6{a}^{\dagger}_7{a}^{\dagger}_4 \nonumber \\
&\qquad \qquad  +{a}^{\dagger}_5{a}^{\dagger}_2{a}^{\dagger}_3{a}^{\dagger}_8  +{a}^{\dagger}_5{a}^{\dagger}_6{a}^{\dagger}_7{a}^{\dagger}_8) \nonumber\\
& \qquad+ \mathcal O(r^{3}) ] \ket{0}
\end{align}
We can then post-select, by measuring all 4 individual Casimir operators (\fig{SpinMeasure}) only the terms that create 4 Qbits, i.e., create {\em exactly one photon, either in 1 or in 5}, and the same for (26), (37), and (48). This yields
\begin{widetext}
\begin{align}
\ket{\psi_f} =&  [G_{12}G_{34} ({a}^{\dagger}_1{a}^{\dagger}_2{a}^{\dagger}_3{a}^{\dagger}_4 + {a}^{\dagger}_1{a}^{\dagger}_2{a}^{\dagger}_7{a}^{\dagger}_8 
+{a}^{\dagger}_5{a}^{\dagger}_6{a}^{\dagger}_3{a}^{\dagger}_4 +{a}^{\dagger}_5{a}^{\dagger}_6{a}^{\dagger}_7{a}^{\dagger}_8) \nonumber \\  
& +G_{23}G_{14} ({a}^{\dagger}_1{a}^{\dagger}_2{a}^{\dagger}_3{a}^{\dagger}_4
+{a}^{\dagger}_1{a}^{\dagger}_6{a}^{\dagger}_7{a}^{\dagger}_4 
+{a}^{\dagger}_5{a}^{\dagger}_2{a}^{\dagger}_3{a}^{\dagger}_8  +{a}^{\dagger}_5{a}^{\dagger}_6{a}^{\dagger}_7{a}^{\dagger}_8)]\ket{0} \nonumber \\
=& a\Bigl(|\uparrow\rangle_{15}|\uparrow\rangle_{26}|\downarrow\rangle_{37}|\downarrow\rangle_{48}
+|\downarrow\rangle_{15}|\downarrow\rangle_{26}|\uparrow\rangle_{37}|\uparrow\rangle_{48}\Bigr)  
+b\Bigl(|\uparrow\rangle_{15}|\downarrow\rangle_{26}|\downarrow\rangle_{37}|\uparrow\rangle_{48}+|\downarrow\rangle_{15}|\uparrow\rangle_{26}|\uparrow\rangle_{37}|\downarrow\rangle_{48}\Bigr) \nonumber \\
&  +(a+b)\Bigl(|\uparrow\rangle_{15}|\uparrow\rangle_{26}|\uparrow\rangle_{37}|\uparrow\rangle_{48}+|\downarrow\rangle_{15}|\downarrow\rangle_{26}|\downarrow\rangle_{37}|\downarrow\rangle_{48}\Bigr) \qquad 
\end{align}
\end{widetext}
where $G_{12}G_{34}=a$ and $ G_{23}G_{14}=b$. We now make the same single-mode unitary operations as mentioned in the last section, considering spins 62 and 84 instead of 26 and 48 and phase-shifting Qmodes 6 and 8 by $\pi$, and get 
\begin{widetext}
\begin{align}
\ket{\psi_f} &= a\Bigl(|\uparrow\rangle_{15}|\downarrow\rangle_{26}|\downarrow\rangle_{37}|\uparrow\rangle_{48}
+|\downarrow\rangle_{15}|\uparrow\rangle_{26}|\uparrow\rangle_{37}|\downarrow\rangle_{48}\Bigr)   
+b\Bigl(|\uparrow\rangle_{15}|\uparrow\rangle_{26}|\downarrow\rangle_{37}|\downarrow\rangle_{48}+|\downarrow\rangle_{15}|\downarrow\rangle_{26}|\uparrow\rangle_{37}|\uparrow\rangle_{48}\Bigr) \nonumber \\
&  -(a+b)\Bigl(|\uparrow\rangle_{15}|\downarrow\rangle_{26}|\uparrow\rangle_{37}|\downarrow\rangle_{48}+|\downarrow\rangle_{15}|\uparrow\rangle_{26}|\downarrow\rangle_{37}|\uparrow\rangle_{48}\Bigr). 
\end{align}
\end{widetext}
We can check that this state satisfies all 4 nullifiers \eqs{j0}{Jz} and is therefore a total spin zero state. We can of course apply an overall normalization constraint satisfying $2|a|^2+2|b|^2+2|a+b|^2=1$. For different instances of G, we will get completely different states which may or may not be entangled spins.

\subsection{Entanglement characterization}

There is no commonly agreed metric of multipartite entanglement even for Qbits although many candidates exist. Average on Neumann entropy, partial positive trace method, Schmidt decomposition are a few of the possible entanglement metrics. There have been several attempts to compare the various entanglement metrics and to characterize and find maximally entangled states~\cite{Brierley2007}. We will not attempt to compare the degree of entanglement of the states considered and will simply observe whether they are factorisable or not.

It is interesting to explore the influence of the different flavors of \eq{Hgeneral} on the created spin state. 

To begin with, in the square cluster Hamiltonian $H^{(1)}$ of \eq{h1}, we have $-G_{14} =G_{12} =G_{23} =G_{34}$, which implies $a=-b$ and the state becomes 
\begin{align}
|\psi_{f}^{(1)}\rangle 
&= \Bigl(|\uparrow\rangle_{15}|\downarrow\rangle_{37} - |\downarrow\rangle_{15}|\uparrow\rangle_{37} \Bigr) \nonumber\\
& \otimes\Bigl(|\uparrow\rangle_{26}|\downarrow\rangle_{48}-|\downarrow\rangle_{26}|\uparrow\rangle_{48}\Bigr),  
\end{align}  
which is a product state of two spin-0 Bell pairs of Qbits. This state is pairwise entangled but not quadripartite entangled. Therefore Schwinger-pairing 2 quadripartite entangled Qmode square cluster states does not necessarily give us a quadripartite spin entangled state. 

However, the situation changes when we consider Hamiltonian $H^{(2)}$ of \eq{h2}, which does not itself make a CV cluster state. There we have  $G_{14} =G_{12} =G_{23} =G_{34}$, which implies $a=b$, and 
\begin{widetext}
\begin{align}
|\psi_{f}^{(2)}\rangle &= \frac{1}{2\sqrt{3}}\Bigl[|\uparrow\rangle_{15}|\uparrow\rangle_{26}|\downarrow\rangle_{37}|\downarrow\rangle_{48}
+|\downarrow\rangle_{15}|\downarrow\rangle_{26}|\uparrow\rangle_{37}|\uparrow\rangle_{48}  
+|\uparrow\rangle_{15}|\downarrow\rangle_{26}|\uparrow\rangle_{37}|\downarrow\rangle_{48} + |\downarrow\rangle_{15}|\uparrow\rangle_{26}|\downarrow\rangle_{37}|\uparrow\rangle_{48} \nonumber \\
&\qquad\qquad  -2\Bigl(|\uparrow\rangle_{15}|\downarrow\rangle_{26}|\downarrow\rangle_{37}|\uparrow\rangle_{48}+|\downarrow\rangle_{15}|\uparrow\rangle_{26}|\uparrow\rangle_{37}|\downarrow\rangle_{48}\Bigr)\Bigr],
\end{align}
which is a quadripartite entangled state. We notice that this state closely resembles the well-known Dicke state~\cite{Dicke1954}
\begin{align}
|D\rangle &= \frac{1}{\sqrt{6}}\Bigl(|\uparrow\rangle_{15}|\uparrow\rangle_{26}|\downarrow\rangle_{37}|\downarrow\rangle_{48}
+|\downarrow\rangle_{15}|\downarrow\rangle_{26}|\uparrow\rangle_{37}|\uparrow\rangle_{48}  
+|\uparrow\rangle_{15}|\downarrow\rangle_{26}|\uparrow\rangle_{37}|\downarrow\rangle_{48} + |\downarrow\rangle_{15}|\uparrow\rangle_{26}|\downarrow\rangle_{37}|\uparrow\rangle_{48} \nonumber \\ 
&\qquad  \quad
+|\uparrow\rangle_{15}|\downarrow\rangle_{26}|\downarrow\rangle_{37}|\uparrow\rangle_{48}+|\downarrow\rangle_{15}|\uparrow\rangle_{26}|\uparrow\rangle_{37}|\downarrow\rangle_{48}\Bigr)
\end{align}
\end{widetext}
and shares many of its properties. Indeed, while the Dicke state is $|j=2,m=0\rangle$ for the total spin, this state is $|j=0,m=0\rangle$. Moreover, projecting 1 spin in the $J_z$ basis, gives us an entangled state, similar to Dicke states. Projecting then another spin results in a bipartite entangled state with the probability 2/3. Therefore $|\psi_{f}^{(2)}\rangle$ can be used as a open destination teleportation resource and is robust under single Qbit decoherence since it retains some entanglement between the remaining Qbits under such projections. 

\section{Conclusion}
The close connection of twin two-Qmode Gaussian entanglement to maximal bipartite spin entanglement, \eqs{twinEPR}{twinspins}, initially suggested that the simulation of entangled spins using optical Qmodes might be possible. Such a correspondence of Gaussian and nonGaussian Wigner functions, coupled to the availability of photon-number-resolving detection methods, is a fascinating prospect, not to mention its possible implications for quantum simulation.

 We attempted to explore this correspondence by recasting different families of H-graph states as spins, by way of the Schwinger representation. While we found that this correspondence is not straightforward for multipartite systems---and even seems to fail systematically in the case of paired CVGHZ states---we have nonetheless obtained nontrivial results, including a genuine multipartite entangled spin-1/2 state generated by a multimode squeezed, albeit not multipartite entangled CV state. Moreover, closely related Qmode Hamiltonians, which make significantly different states,  can still have the same spin nullifiers, which hints at a possible degeneracy of the nullifier picture in this case.

These results are intriguing and we believe that a possible avenue to better understand the underlying theory might be a general group theoretical approach involving the connections between $\rm Sp(4,\mathbb R)$ and SU(2). This will be the focus of an upcoming extension of this work, along with the investigation of larger spin values.

We are grateful to Natasha Gabay and Nicolas Menicucci at the University of Sydney for nonlocal stimulating discussions and free-flowing exchange of ideas. This work was supported by the U.S. National Science Foundation grants No.\ PHY-1206029, No.\ PHY-0960047, and No.\ PHY-0855632. 


\begin{thebibliography}{38}%
\makeatletter
\providecommand \@ifxundefined [1]{%
 \@ifx{#1\undefined}
}%
\providecommand \@ifnum [1]{%
 \ifnum #1\expandafter \@firstoftwo
 \else \expandafter \@secondoftwo
 \fi
}%
\providecommand \@ifx [1]{%
 \ifx #1\expandafter \@firstoftwo
 \else \expandafter \@secondoftwo
 \fi
}%
\providecommand \natexlab [1]{#1}%
\providecommand \enquote  [1]{``#1''}%
\providecommand \bibnamefont  [1]{#1}%
\providecommand \bibfnamefont [1]{#1}%
\providecommand \citenamefont [1]{#1}%
\providecommand \href@noop [0]{\@secondoftwo}%
\providecommand \href [0]{\begingroup \@sanitize@url \@href}%
\providecommand \@href[1]{\@@startlink{#1}\@@href}%
\providecommand \@@href[1]{\endgroup#1\@@endlink}%
\providecommand \@sanitize@url [0]{\catcode `\\12\catcode `\$12\catcode
  `\&12\catcode `\#12\catcode `\^12\catcode `\_12\catcode `\%12\relax}%
\providecommand \@@startlink[1]{}%
\providecommand \@@endlink[0]{}%
\providecommand \url  [0]{\begingroup\@sanitize@url \@url }%
\providecommand \@url [1]{\endgroup\@href {#1}{\urlprefix }}%
\providecommand \urlprefix  [0]{URL }%
\providecommand \Eprint [0]{\href }%
\providecommand \doibase [0]{http://dx.doi.org/}%
\providecommand \selectlanguage [0]{\@gobble}%
\providecommand \bibinfo  [0]{\@secondoftwo}%
\providecommand \bibfield  [0]{\@secondoftwo}%
\providecommand \translation [1]{[#1]}%
\providecommand \BibitemOpen [0]{}%
\providecommand \bibitemStop [0]{}%
\providecommand \bibitemNoStop [0]{.\EOS\space}%
\providecommand \EOS [0]{\spacefactor3000\relax}%
\providecommand \BibitemShut  [1]{\csname bibitem#1\endcsname}%
\let\auto@bib@innerbib\@empty
\bibitem [{\citenamefont {Pfister}\ \emph {et~al.}(2004)\citenamefont
  {Pfister}, \citenamefont {Feng}, \citenamefont {Jennings}, \citenamefont
  {Pooser},\ and\ \citenamefont {Xie}}]{Pfister2004}%
  \BibitemOpen
  \bibfield  {author} {\bibinfo {author} {\bibfnamefont {O.}~\bibnamefont
  {Pfister}}, \bibinfo {author} {\bibfnamefont {S.}~\bibnamefont {Feng}},
  \bibinfo {author} {\bibfnamefont {G.}~\bibnamefont {Jennings}}, \bibinfo
  {author} {\bibfnamefont {R.}~\bibnamefont {Pooser}}, \ and\ \bibinfo {author}
  {\bibfnamefont {D.}~\bibnamefont {Xie}},\ }\href {\doibase
  10.1103/PhysRevA.70.020302} {\bibfield  {journal} {\bibinfo  {journal}
  {Phys.\ Rev.\ A}\ }\textbf {\bibinfo {volume} {70}},\ \bibinfo {pages}
  {020302} (\bibinfo {year} {2004})}\BibitemShut {NoStop}%
\bibitem [{\citenamefont {Menicucci}\ \emph {et~al.}(2008)\citenamefont
  {Menicucci}, \citenamefont {Flammia},\ and\ \citenamefont
  {Pfister}}]{Menicucci2008}%
  \BibitemOpen
  \bibfield  {author} {\bibinfo {author} {\bibfnamefont {N.~C.}\ \bibnamefont
  {Menicucci}}, \bibinfo {author} {\bibfnamefont {S.~T.}\ \bibnamefont
  {Flammia}}, \ and\ \bibinfo {author} {\bibfnamefont {O.}~\bibnamefont
  {Pfister}},\ }\href {\doibase 10.1103/PhysRevLett.101.130501} {\bibfield
  {journal} {\bibinfo  {journal} {Phys.\ Rev.\ Lett.}\ }\textbf {\bibinfo
  {volume} {101}},\ \bibinfo {pages} {130501} (\bibinfo {year}
  {2008})}\BibitemShut {NoStop}%
\bibitem [{\citenamefont {Pysher}\ \emph {et~al.}(2011)\citenamefont {Pysher},
  \citenamefont {Miwa}, \citenamefont {Shahrokhshahi}, \citenamefont
  {Bloomer},\ and\ \citenamefont {Pfister}}]{Pysher2011}%
  \BibitemOpen
  \bibfield  {author} {\bibinfo {author} {\bibfnamefont {M.}~\bibnamefont
  {Pysher}}, \bibinfo {author} {\bibfnamefont {Y.}~\bibnamefont {Miwa}},
  \bibinfo {author} {\bibfnamefont {R.}~\bibnamefont {Shahrokhshahi}}, \bibinfo
  {author} {\bibfnamefont {R.}~\bibnamefont {Bloomer}}, \ and\ \bibinfo
  {author} {\bibfnamefont {O.}~\bibnamefont {Pfister}},\ }\href {\doibase
  10.1103/PhysRevLett.107.030505} {\bibfield  {journal} {\bibinfo  {journal}
  {Phys.\ Rev.\ Lett.}\ }\textbf {\bibinfo {volume} {107}},\ \bibinfo {pages}
  {030505} (\bibinfo {year} {2011})}\BibitemShut {NoStop}%
\bibitem [{\citenamefont {Yokoyama}\ \emph {et~al.}(2013)\citenamefont
  {Yokoyama}, \citenamefont {Ukai}, \citenamefont {Armstrong}, \citenamefont
  {Sornphiphatphong}, \citenamefont {Kaji}, \citenamefont {Suzuki},
  \citenamefont {ichi Yoshikawa}, \citenamefont {Yonezawa}, \citenamefont
  {Menicucci},\ and\ \citenamefont {Furusawa}}]{Yokoyama2013}%
  \BibitemOpen
  \bibfield  {author} {\bibinfo {author} {\bibfnamefont {S.}~\bibnamefont
  {Yokoyama}}, \bibinfo {author} {\bibfnamefont {R.}~\bibnamefont {Ukai}},
  \bibinfo {author} {\bibfnamefont {S.~C.}\ \bibnamefont {Armstrong}}, \bibinfo
  {author} {\bibfnamefont {C.}~\bibnamefont {Sornphiphatphong}}, \bibinfo
  {author} {\bibfnamefont {T.}~\bibnamefont {Kaji}}, \bibinfo {author}
  {\bibfnamefont {S.}~\bibnamefont {Suzuki}}, \bibinfo {author} {\bibfnamefont
  {J.}~\bibnamefont {ichi Yoshikawa}}, \bibinfo {author} {\bibfnamefont
  {H.}~\bibnamefont {Yonezawa}}, \bibinfo {author} {\bibfnamefont {N.~C.}\
  \bibnamefont {Menicucci}}, \ and\ \bibinfo {author} {\bibfnamefont
  {A.}~\bibnamefont {Furusawa}},\ }\href@noop {} {\bibfield  {journal}
  {\bibinfo  {journal} {arXiv:1306.3366 [quant-ph]}\ } (\bibinfo {year}
  {2013})}\BibitemShut {NoStop}%
\bibitem [{\citenamefont {Roslund}\ \emph {et~al.}(2013)\citenamefont
  {Roslund}, \citenamefont {{Medeiros de Ara\'ujo}}, \citenamefont {Jiang},
  \citenamefont {Fabre},\ and\ \citenamefont {Treps}}]{Roslund2013}%
  \BibitemOpen
  \bibfield  {author} {\bibinfo {author} {\bibfnamefont {J.}~\bibnamefont
  {Roslund}}, \bibinfo {author} {\bibfnamefont {R.}~\bibnamefont {{Medeiros de
  Ara\'ujo}}}, \bibinfo {author} {\bibfnamefont {S.}~\bibnamefont {Jiang}},
  \bibinfo {author} {\bibfnamefont {C.}~\bibnamefont {Fabre}}, \ and\ \bibinfo
  {author} {\bibfnamefont {N.}~\bibnamefont {Treps}},\ }\href@noop {}
  {\bibfield  {journal} {\bibinfo  {journal} {arXiv:1307.1216 [quant-ph]}\ }
  (\bibinfo {year} {2013})}\BibitemShut {NoStop}%
\bibitem [{\citenamefont {Flammia}\ \emph {et~al.}(2009)\citenamefont
  {Flammia}, \citenamefont {Menicucci},\ and\ \citenamefont
  {Pfister}}]{Flammia2009}%
  \BibitemOpen
  \bibfield  {author} {\bibinfo {author} {\bibfnamefont {S.~T.}\ \bibnamefont
  {Flammia}}, \bibinfo {author} {\bibfnamefont {N.~C.}\ \bibnamefont
  {Menicucci}}, \ and\ \bibinfo {author} {\bibfnamefont {O.}~\bibnamefont
  {Pfister}},\ }\href@noop {} {\bibfield  {journal} {\bibinfo  {journal} {J.
  Phys.\ B,}\ }\textbf {\bibinfo {volume} {42}},\ \bibinfo {pages} {114009}
  (\bibinfo {year} {2009})}\BibitemShut {NoStop}%
\bibitem [{\citenamefont {Bell}(1987)}]{Bell1987}%
  \BibitemOpen
  \bibfield  {author} {\bibinfo {author} {\bibfnamefont {J.~S.}\ \bibnamefont
  {Bell}},\ }\enquote {\bibinfo {title} {Speakable and unspeakable in quantum
  mechanics},}\ \ (\bibinfo  {publisher} {Cambridge University Press},\
  \bibinfo {year} {1987})\ Chap.\ \bibinfo {chapter} {21, {\em ``EPR
  correlations and EPW distributions''}}, pp.\ \bibinfo {pages}
  {196--200}\BibitemShut {NoStop}%
\bibitem [{\citenamefont {Eisert}\ \emph {et~al.}(2002)\citenamefont {Eisert},
  \citenamefont {Scheel},\ and\ \citenamefont {Plenio}}]{Eisert2002}%
  \BibitemOpen
  \bibfield  {author} {\bibinfo {author} {\bibfnamefont {J.}~\bibnamefont
  {Eisert}}, \bibinfo {author} {\bibfnamefont {S.}~\bibnamefont {Scheel}}, \
  and\ \bibinfo {author} {\bibfnamefont {M.~B.}\ \bibnamefont {Plenio}},\
  }\href {\doibase 10.1103/PhysRevLett.89.137903} {\bibfield  {journal}
  {\bibinfo  {journal} {Phys.\ Rev.\ Lett.}\ }\textbf {\bibinfo {volume}
  {89}},\ \bibinfo {pages} {137903} (\bibinfo {year} {2002})}\BibitemShut
  {NoStop}%
\bibitem [{\citenamefont {Niset}\ \emph {et~al.}(2009)\citenamefont {Niset},
  \citenamefont {{Fiur\'a\v sek}},\ and\ \citenamefont {Cerf}}]{Niset2009}%
  \BibitemOpen
  \bibfield  {author} {\bibinfo {author} {\bibfnamefont {J.}~\bibnamefont
  {Niset}}, \bibinfo {author} {\bibfnamefont {J.}~\bibnamefont {{Fiur\'a\v
  sek}}}, \ and\ \bibinfo {author} {\bibfnamefont {N.~J.}\ \bibnamefont
  {Cerf}},\ }\href@noop {} {\bibfield  {journal} {\bibinfo  {journal} {Phys.\
  Rev.\ Lett.}\ }\textbf {\bibinfo {volume} {102}},\ \bibinfo {pages} {120501}
  (\bibinfo {year} {2009})}\BibitemShut {NoStop}%
\bibitem [{\citenamefont {Menicucci}\ \emph {et~al.}(2006)\citenamefont
  {Menicucci}, \citenamefont {{van Loock}}, \citenamefont {Gu}, \citenamefont
  {Weedbrook}, \citenamefont {Ralph},\ and\ \citenamefont
  {Nielsen}}]{Menicucci2006}%
  \BibitemOpen
  \bibfield  {author} {\bibinfo {author} {\bibfnamefont {N.~C.}\ \bibnamefont
  {Menicucci}}, \bibinfo {author} {\bibfnamefont {P.}~\bibnamefont {{van
  Loock}}}, \bibinfo {author} {\bibfnamefont {M.}~\bibnamefont {Gu}}, \bibinfo
  {author} {\bibfnamefont {C.}~\bibnamefont {Weedbrook}}, \bibinfo {author}
  {\bibfnamefont {T.~C.}\ \bibnamefont {Ralph}}, \ and\ \bibinfo {author}
  {\bibfnamefont {M.~A.}\ \bibnamefont {Nielsen}},\ }\href {\doibase
  doi:10.1103/PhysRevLett.97.110501} {\bibfield  {journal} {\bibinfo  {journal}
  {Phys.\ Rev.\ Lett.}\ }\textbf {\bibinfo {volume} {97}},\ \bibinfo {pages}
  {110501} (\bibinfo {year} {2006})}\BibitemShut {NoStop}%
\bibitem [{\citenamefont {Schwinger}(1965)}]{Schwinger1965}%
  \BibitemOpen
  \bibfield  {author} {\bibinfo {author} {\bibfnamefont {J.}~\bibnamefont
  {Schwinger}},\ }\enquote {\bibinfo {title} {Quantum theory of angular
  momentum},}\ \ (\bibinfo  {publisher} {Academic Press},\ \bibinfo {year}
  {1965})\ Chap.~\bibinfo {chapter} {{\em ``On angular momentum''}}, pp.\
  \bibinfo {pages} {229--279}\BibitemShut {NoStop}%
\bibitem [{\citenamefont {Lita}\ \emph {et~al.}(2008)\citenamefont {Lita},
  \citenamefont {Miller},\ and\ \citenamefont {Nam}}]{Lita2008}%
  \BibitemOpen
  \bibfield  {author} {\bibinfo {author} {\bibfnamefont {A.~E.}\ \bibnamefont
  {Lita}}, \bibinfo {author} {\bibfnamefont {A.~J.}\ \bibnamefont {Miller}}, \
  and\ \bibinfo {author} {\bibfnamefont {S.~W.}\ \bibnamefont {Nam}},\
  }\href@noop {} {\bibfield  {journal} {\bibinfo  {journal} {Opt.\ Expr.}\
  }\textbf {\bibinfo {volume} {16}},\ \bibinfo {pages} {3032} (\bibinfo {year}
  {2008})}\BibitemShut {NoStop}%
\bibitem [{\citenamefont {Hudson}(1974)}]{Hudson1974}%
  \BibitemOpen
  \bibfield  {author} {\bibinfo {author} {\bibfnamefont {R.~L.}\ \bibnamefont
  {Hudson}},\ }\href {\doibase DOI: 10.1016/0034-4877(74)90007-X} {\bibfield
  {journal} {\bibinfo  {journal} {Rep.\ Math.\ Phys.}\ }\textbf {\bibinfo
  {volume} {6}},\ \bibinfo {pages} {249} (\bibinfo {year} {1974})}\BibitemShut
  {NoStop}%
\bibitem [{\citenamefont {Feynman}(1982)}]{Feynman1982}%
  \BibitemOpen
  \bibfield  {author} {\bibinfo {author} {\bibfnamefont {R.~P.}\ \bibnamefont
  {Feynman}},\ }\href@noop {} {\bibfield  {journal} {\bibinfo  {journal} {Int.
  J. Theor. Phys.}\ }\textbf {\bibinfo {volume} {21}},\ \bibinfo {pages} {467}
  (\bibinfo {year} {1982})}\BibitemShut {NoStop}%
\bibitem [{\citenamefont {Mermin}(2003)}]{Mermin2003}%
  \BibitemOpen
  \bibfield  {author} {\bibinfo {author} {\bibfnamefont {N.~D.}\ \bibnamefont
  {Mermin}},\ }\href {\doibase 10.1119/1.1522741} {\bibfield  {journal}
  {\bibinfo  {journal} {American Journal of Physics}\ }\textbf {\bibinfo
  {volume} {71}},\ \bibinfo {pages} {23} (\bibinfo {year} {2003})}\BibitemShut
  {NoStop}%
\bibitem [{\citenamefont {Gabay}\ and\ \citenamefont
  {Menicucci}(2013)}]{Gabay2013}%
  \BibitemOpen
  \bibfield  {author} {\bibinfo {author} {\bibfnamefont {N.~C.}\ \bibnamefont
  {Gabay}}\ and\ \bibinfo {author} {\bibfnamefont {N.~C.}\ \bibnamefont
  {Menicucci}},\ }\href@noop {} {\bibfield  {journal} {\bibinfo  {journal} {to
  be released simultaneously}\ } (\bibinfo {year} {2013})}\BibitemShut
  {NoStop}%
\bibitem [{\citenamefont {Menicucci}\ \emph {et~al.}(2011)\citenamefont
  {Menicucci}, \citenamefont {Flammia},\ and\ \citenamefont {van
  Loock}}]{Menicucci2011}%
  \BibitemOpen
  \bibfield  {author} {\bibinfo {author} {\bibfnamefont {N.~C.}\ \bibnamefont
  {Menicucci}}, \bibinfo {author} {\bibfnamefont {S.~T.}\ \bibnamefont
  {Flammia}}, \ and\ \bibinfo {author} {\bibfnamefont {P.}~\bibnamefont {van
  Loock}},\ }\href@noop {} {\bibfield  {journal} {\bibinfo  {journal} {Phys.\
  Rev.\ A}\ }\textbf {\bibinfo {volume} {83}},\ \bibinfo {pages} {042335}
  (\bibinfo {year} {2011})}\BibitemShut {NoStop}%
\bibitem [{\citenamefont {Menicucci}(2011)}]{Menicucci2011a}%
  \BibitemOpen
  \bibfield  {author} {\bibinfo {author} {\bibfnamefont {N.~C.}\ \bibnamefont
  {Menicucci}},\ }\href {\doibase 10.1103/PhysRevA.83.062314} {\bibfield
  {journal} {\bibinfo  {journal} {Phys.\ Rev.\ A}\ }\textbf {\bibinfo {volume}
  {83}},\ \bibinfo {pages} {062314} (\bibinfo {year} {2011})}\BibitemShut
  {NoStop}%
\bibitem [{\citenamefont {Yurke}\ \emph {et~al.}(1986)\citenamefont {Yurke},
  \citenamefont {McCall},\ and\ \citenamefont {Klauder}}]{Yurke1986a}%
  \BibitemOpen
  \bibfield  {author} {\bibinfo {author} {\bibfnamefont {B.}~\bibnamefont
  {Yurke}}, \bibinfo {author} {\bibfnamefont {S.}~\bibnamefont {McCall}}, \
  and\ \bibinfo {author} {\bibfnamefont {J.}~\bibnamefont {Klauder}},\
  }\href@noop {} {\bibfield  {journal} {\bibinfo  {journal} {Phys.\ Rev.\ A}\
  }\textbf {\bibinfo {volume} {33}},\ \bibinfo {pages} {4033} (\bibinfo {year}
  {1986})}\BibitemShut {NoStop}%
\bibitem [{\citenamefont {Kim}\ \emph {et~al.}(1998)\citenamefont {Kim},
  \citenamefont {Pfister}, \citenamefont {Holland}, \citenamefont {Noh},\ and\
  \citenamefont {Hall}}]{Kim1998}%
  \BibitemOpen
  \bibfield  {author} {\bibinfo {author} {\bibfnamefont {T.}~\bibnamefont
  {Kim}}, \bibinfo {author} {\bibfnamefont {O.}~\bibnamefont {Pfister}},
  \bibinfo {author} {\bibfnamefont {M.~J.}\ \bibnamefont {Holland}}, \bibinfo
  {author} {\bibfnamefont {J.}~\bibnamefont {Noh}}, \ and\ \bibinfo {author}
  {\bibfnamefont {J.~L.}\ \bibnamefont {Hall}},\ }\href@noop {} {\bibfield
  {journal} {\bibinfo  {journal} {Phys.\ Rev.\ A}\ }\textbf {\bibinfo {volume}
  {57}},\ \bibinfo {pages} {4004} (\bibinfo {year} {1998})}\BibitemShut
  {NoStop}%
\bibitem [{\citenamefont {Evans}\ and\ \citenamefont
  {Pfister}(2011)}]{Evans2011}%
  \BibitemOpen
  \bibfield  {author} {\bibinfo {author} {\bibfnamefont {R.}~\bibnamefont
  {Evans}}\ and\ \bibinfo {author} {\bibfnamefont {O.}~\bibnamefont
  {Pfister}},\ }\href@noop {} {\bibfield  {journal} {\bibinfo  {journal}
  {Quantum Information and Computation}\ }\textbf {\bibinfo {volume} {11}},\
  \bibinfo {pages} {820} (\bibinfo {year} {2011})}\BibitemShut {NoStop}%
\bibitem [{\citenamefont {Bowen}\ \emph {et~al.}(2002)\citenamefont {Bowen},
  \citenamefont {Schnabel}, \citenamefont {Bachor},\ and\ \citenamefont
  {Lam}}]{Bowen2002}%
  \BibitemOpen
  \bibfield  {author} {\bibinfo {author} {\bibfnamefont {W.}~\bibnamefont
  {Bowen}}, \bibinfo {author} {\bibfnamefont {R.}~\bibnamefont {Schnabel}},
  \bibinfo {author} {\bibfnamefont {H.}~\bibnamefont {Bachor}}, \ and\ \bibinfo
  {author} {\bibfnamefont {P.}~\bibnamefont {Lam}},\ }\href@noop {} {\bibfield
  {journal} {\bibinfo  {journal} {Phys.\ Rev.\ Lett.}\ }\textbf {\bibinfo
  {volume} {88}},\ \bibinfo {pages} {093601} (\bibinfo {year}
  {2002})}\BibitemShut {NoStop}%
\bibitem [{\citenamefont {Drummond}(1983)}]{Drummond1983}%
  \BibitemOpen
  \bibfield  {author} {\bibinfo {author} {\bibfnamefont {P.~D.}\ \bibnamefont
  {Drummond}},\ }\href {\doibase 10.1103/PhysRevLett.50.1407} {\bibfield
  {journal} {\bibinfo  {journal} {Phys.\ Rev.\ Lett.}\ }\textbf {\bibinfo
  {volume} {50}},\ \bibinfo {pages} {1407} (\bibinfo {year}
  {1983})}\BibitemShut {NoStop}%
\bibitem [{\citenamefont {Reid}\ \emph {et~al.}(2002)\citenamefont {Reid},
  \citenamefont {Munro},\ and\ \citenamefont {De~Martini}}]{Reid2002}%
  \BibitemOpen
  \bibfield  {author} {\bibinfo {author} {\bibfnamefont {M.~D.}\ \bibnamefont
  {Reid}}, \bibinfo {author} {\bibfnamefont {W.~J.}\ \bibnamefont {Munro}}, \
  and\ \bibinfo {author} {\bibfnamefont {F.}~\bibnamefont {De~Martini}},\
  }\href {\doibase 10.1103/PhysRevA.66.033801} {\bibfield  {journal} {\bibinfo
  {journal} {Phys.\ Rev.\ A}\ }\textbf {\bibinfo {volume} {66}},\ \bibinfo
  {pages} {033801} (\bibinfo {year} {2002})}\BibitemShut {NoStop}%
\bibitem [{\citenamefont {Mermin}(1980)}]{Mermin1980}%
  \BibitemOpen
  \bibfield  {author} {\bibinfo {author} {\bibfnamefont {N.~D.}\ \bibnamefont
  {Mermin}},\ }\href {\doibase 10.1103/PhysRevD.22.356} {\bibfield  {journal}
  {\bibinfo  {journal} {Phys.\ Rev.\ D}\ }\textbf {\bibinfo {volume} {22}},\
  \bibinfo {pages} {356} (\bibinfo {year} {1980})}\BibitemShut {NoStop}%
\bibitem [{\citenamefont {Gerry}(2001)}]{Gerry2001}%
  \BibitemOpen
  \bibfield  {author} {\bibinfo {author} {\bibfnamefont {C.~C.}\ \bibnamefont
  {Gerry}},\ }\href {http://www.opticsexpress.org/abstract.cfm?URI=oe-8-2-76}
  {\bibfield  {journal} {\bibinfo  {journal} {Opt.\ Express}\ ,\ \bibinfo
  {pages} {76}} (\bibinfo {year} {2001})}\BibitemShut {NoStop}%
\bibitem [{\citenamefont {W\'odkiewicz}\ and\ \citenamefont
  {Eberly}(1985)}]{Wodkiewicz1985}%
  \BibitemOpen
  \bibfield  {author} {\bibinfo {author} {\bibfnamefont {K.}~\bibnamefont
  {W\'odkiewicz}}\ and\ \bibinfo {author} {\bibfnamefont {J.}~\bibnamefont
  {Eberly}},\ }\href@noop {} {\bibfield  {journal} {\bibinfo  {journal} {J.
  Opt.\ Soc.\ Am.\ B}\ }\textbf {\bibinfo {volume} {2}},\ \bibinfo {pages}
  {458} (\bibinfo {year} {1985})}\BibitemShut {NoStop}%
\bibitem [{\citenamefont {Arvind}\ \emph {et~al.}(1995)\citenamefont {Arvind},
  \citenamefont {Dutta}, \citenamefont {Mukunda},\ and\ \citenamefont
  {Simon}}]{Arvind1995}%
  \BibitemOpen
  \bibfield  {author} {\bibinfo {author} {\bibnamefont {Arvind}}, \bibinfo
  {author} {\bibfnamefont {B.}~\bibnamefont {Dutta}}, \bibinfo {author}
  {\bibfnamefont {N.}~\bibnamefont {Mukunda}}, \ and\ \bibinfo {author}
  {\bibfnamefont {R.}~\bibnamefont {Simon}},\ }\href {\doibase
  10.1103/PhysRevA.52.1609} {\bibfield  {journal} {\bibinfo  {journal} {Phys.
  Rev. A}\ }\textbf {\bibinfo {volume} {52}},\ \bibinfo {pages} {1609}
  (\bibinfo {year} {1995})}\BibitemShut {NoStop}%
\bibitem [{\citenamefont {Simon}(2000)}]{Simon2000}%
  \BibitemOpen
  \bibfield  {author} {\bibinfo {author} {\bibfnamefont {R.}~\bibnamefont
  {Simon}},\ }\href@noop {} {\bibfield  {journal} {\bibinfo  {journal} {Phys.\
  Rev.\ Lett.}\ }\textbf {\bibinfo {volume} {84}},\ \bibinfo {pages} {2726}
  (\bibinfo {year} {2000})}\BibitemShut {NoStop}%
\bibitem [{\citenamefont {Zaidi}\ \emph {et~al.}(2008)\citenamefont {Zaidi},
  \citenamefont {Menicucci}, \citenamefont {Flammia}, \citenamefont {Bloomer},
  \citenamefont {Pysher},\ and\ \citenamefont {Pfister}}]{Zaidi2008}%
  \BibitemOpen
  \bibfield  {author} {\bibinfo {author} {\bibfnamefont {H.}~\bibnamefont
  {Zaidi}}, \bibinfo {author} {\bibfnamefont {N.~C.}\ \bibnamefont
  {Menicucci}}, \bibinfo {author} {\bibfnamefont {S.~T.}\ \bibnamefont
  {Flammia}}, \bibinfo {author} {\bibfnamefont {R.}~\bibnamefont {Bloomer}},
  \bibinfo {author} {\bibfnamefont {M.}~\bibnamefont {Pysher}}, \ and\ \bibinfo
  {author} {\bibfnamefont {O.}~\bibnamefont {Pfister}},\ }\href@noop {}
  {\bibfield  {journal} {\bibinfo  {journal} {Laser Phys.}\ }\textbf {\bibinfo
  {volume} {18}},\ \bibinfo {pages} {659} (\bibinfo {year} {2008})},\ \bibinfo
  {note} {revised version at http://arxiv.org/pdf/0710.4980v3}\BibitemShut
  {NoStop}%
\bibitem [{\citenamefont {Menicucci}(2013)}]{Menicucci2013}%
  \BibitemOpen
  \bibfield  {author} {\bibinfo {author} {\bibfnamefont {N.~C.}\ \bibnamefont
  {Menicucci}},\ }\href@noop {} {\bibfield  {journal} {\bibinfo  {journal} {in
  preparation}\ } (\bibinfo {year} {2013})}\BibitemShut {NoStop}%
\bibitem [{\citenamefont {Gu}\ \emph {et~al.}(2009)\citenamefont {Gu},
  \citenamefont {Weedbrook}, \citenamefont {Menicucci}, \citenamefont {Ralph},\
  and\ \citenamefont {van Loock}}]{Gu2009}%
  \BibitemOpen
  \bibfield  {author} {\bibinfo {author} {\bibfnamefont {M.}~\bibnamefont
  {Gu}}, \bibinfo {author} {\bibfnamefont {C.}~\bibnamefont {Weedbrook}},
  \bibinfo {author} {\bibfnamefont {N.~C.}\ \bibnamefont {Menicucci}}, \bibinfo
  {author} {\bibfnamefont {T.~C.}\ \bibnamefont {Ralph}}, \ and\ \bibinfo
  {author} {\bibfnamefont {P.}~\bibnamefont {van Loock}},\ }\href@noop {}
  {\bibfield  {journal} {\bibinfo  {journal} {Phys.\ Rev.\ A}\ }\textbf
  {\bibinfo {volume} {79}},\ \bibinfo {pages} {062318} (\bibinfo {year}
  {2009})}\BibitemShut {NoStop}%
\bibitem [{\citenamefont {Hyllus}\ and\ \citenamefont
  {Eisert}(2006)}]{Hyllus2006}%
  \BibitemOpen
  \bibfield  {author} {\bibinfo {author} {\bibfnamefont {P.}~\bibnamefont
  {Hyllus}}\ and\ \bibinfo {author} {\bibfnamefont {J.}~\bibnamefont
  {Eisert}},\ }\href {http://stacks.iop.org/1367-2630/8/i=4/a=051} {\bibfield
  {journal} {\bibinfo  {journal} {New J. Phys.}\ }\textbf {\bibinfo {volume}
  {8}},\ \bibinfo {pages} {51} (\bibinfo {year} {2006})}\BibitemShut {NoStop}%
\bibitem [{\citenamefont {Gottesman}(1997)}]{Gottesman1997}%
  \BibitemOpen
  \bibfield  {author} {\bibinfo {author} {\bibfnamefont {D.}~\bibnamefont
  {Gottesman}},\ }\emph {\bibinfo {title} {Stabilizer codes and quantum error
  correction}},\ \href@noop {} {Ph.D. thesis},\ \bibinfo  {school} {California
  Institute of Technology}, \bibinfo {address} {Pasadena, CA} (\bibinfo {year}
  {1997}),\ \Eprint {http://arxiv.org/abs/quant-ph/9705052} {quant-ph/9705052}
  \BibitemShut {NoStop}%
\bibitem [{\citenamefont {{van Loock}}\ and\ \citenamefont
  {Braunstein}(2003)}]{Braunstein2003a}%
  \BibitemOpen
  \bibfield  {author} {\bibinfo {author} {\bibfnamefont {P.}~\bibnamefont {{van
  Loock}}}\ and\ \bibinfo {author} {\bibfnamefont {S.~L.}\ \bibnamefont
  {Braunstein}},\ }\enquote {\bibinfo {title} {Quantum information with
  continuous variables},}\ \ (\bibinfo  {publisher} {Kluwer Academic},\
  \bibinfo {year} {2003})\ Chap.\ \bibinfo {chapter} {Multipartite entanglement
  for continuous variables}\BibitemShut {NoStop}%
\bibitem [{\citenamefont {Zhang}\ and\ \citenamefont
  {Braunstein}(2006)}]{Zhang2006}%
  \BibitemOpen
  \bibfield  {author} {\bibinfo {author} {\bibfnamefont {J.}~\bibnamefont
  {Zhang}}\ and\ \bibinfo {author} {\bibfnamefont {S.~L.}\ \bibnamefont
  {Braunstein}},\ }\href@noop {} {\bibfield  {journal} {\bibinfo  {journal}
  {Phys.\ Rev.\ A}\ }\textbf {\bibinfo {volume} {73}},\ \bibinfo {eid} {032318}
  (\bibinfo {year} {2006})}\BibitemShut {NoStop}%
\bibitem [{\citenamefont {Brierley}\ and\ \citenamefont
  {Higuchi}(2007)}]{Brierley2007}%
  \BibitemOpen
  \bibfield  {author} {\bibinfo {author} {\bibfnamefont {S.}~\bibnamefont
  {Brierley}}\ and\ \bibinfo {author} {\bibfnamefont {A.}~\bibnamefont
  {Higuchi}},\ }\href@noop {} {\bibfield  {journal} {\bibinfo  {journal} {J.
  Phys.\ A: Math.\ Theor.}\ }\textbf {\bibinfo {volume} {40}},\ \bibinfo
  {pages} {8455} (\bibinfo {year} {2007})}\BibitemShut {NoStop}%
\bibitem [{\citenamefont {Dicke}(1954)}]{Dicke1954}%
  \BibitemOpen
  \bibfield  {author} {\bibinfo {author} {\bibfnamefont {R.}~\bibnamefont
  {Dicke}},\ }\href@noop {} {\bibfield  {journal} {\bibinfo  {journal} {Phys.\
  Rev.}\ }\textbf {\bibinfo {volume} {93}},\ \bibinfo {pages} {99} (\bibinfo
  {year} {1954})}\BibitemShut {NoStop}%
\end{thebibliography}

%

\end{document}